\documentclass[authoryear,12pt]{elsarticle}      
\usepackage[english]{babel}
\usepackage[toc,page]{appendix}
\usepackage{amssymb}        
\usepackage{amsthm}         
\usepackage{amsmath}
\usepackage{natbib}         
\usepackage{graphicx}       
\usepackage{subfigure}      
\usepackage{color}
\usepackage{gensymb} 

\journal{Journal of the Mechanics and Physics of Solids}

\newcommand{\tensor}[1]{\boldsymbol{#1}}
\newcommand{\ftensor}[1]{\mathbb{#1}}

\newcommand{\rd}{\mathrm{d}}

\begin{document}

\begin{frontmatter}

\title{Discrete dislocation dynamics simulations of dislocation-$\theta'$ precipitate interaction in Al-Cu alloys} 

\author{R. Santos-G\"uemes$^{1, 2}$}
\author{G. Esteban-Manzanares$^{1, 2}$}
\author{I. Papadimitriou$^1$}
\author{J. Segurado$^{1, 2}$}
\author{L. Capolungo$^{3}$}
\author{J. LLorca$^{1, 2, }$\corref{cor1}}
\address{$^1$ IMDEA Materials Institute, C/ Eric Kandel 2, 28906, Getafe, Madrid, Spain. \\  $^2$ Department of Materials Science, Polytechnic University of Madrid/Universidad Polit\'ecnica de Madrid, E. T. S. de Ingenieros de Caminos. 28040 - Madrid, Spain. \\  $^3$ Material Science and Technology Division, MST-8, Los Alamos National Laboratory, Los Alamos 87545 NM, USA.}

\cortext[cor1]{Corresponding author}

\begin{abstract}

The mechanisms of dislocation/precipitate interaction were studied by means of discrete dislocation dynamics within a multiscale approach. Simulations were carried out using the discrete continuous method in combination with a fast Fourier transform solver to compute the mechanical fields \citep{Capolungo2015_DDDFFT}. The original simulation strategy was modified to include straight dislocation segments by means of the field dislocation mechanics method and was applied to simulate the interaction of an edge dislocation with a $\theta'$ precipitate in an Al-Cu alloy.  It was found that the elastic mismatch has a negligible influence on the dislocation/precipitate interaction in the Al-Cu system. Moreover, the influence of the precipitate aspect ratio and orientation was reasonably well captured by the simple Orowan model in the absence of the stress-free transformation strain. Nevertheless, the introduction of the stress-free transformation strain led to dramatic changes in the dislocation/precipitate interaction and in the critical resolved shear stress to overcome the precipitate, particularly in the case of precipitates with small aspect ratio. The new multiscale approach to study the dislocation/precipitate interactions opens the possibility to obtain quantitative estimations of the strengthening provided by precipitates in metallic alloys taking into account the microstructural details.

\end{abstract}


\end{frontmatter}

\section{Introduction}
\label{sec:intro}

Plastic deformation in metallic alloys is carried by dislocation slip and strengthening is achieved with obstacles that hinder the motion of dislocations. These obstacles can take the form of dislocations during deformation, solute atoms in the lattice, grain boundaries, etc. but precipitation hardening is well established as the most effective mechanism  to increase the yield strength in metallic alloys \citep{ardell1985precipitation}. Obviously, the strengthening provided by the dispersion of second phases depends on the chemical composition, size, shape, orientation, spatial distribution, etc. of the precipitates, which have been optimized over the years through costly, experimental trial-and-error approaches. Nevertheless, these strategies are being overcome by recent advances in multiscale modelling approaches based on the coupling of {\it ab initio} and atomistic simulations with computational thermodynamics and phase-field models that allow an accurate prediction of the precipitate features as a function of the alloy chemical composition and thermo-mechanical treatment \citep{LGL13, LBL17, JIH14}.

In addition to these tools, the design of metallic alloys in which precipitate strengthening has been maximized 
requires the development of multiscale modelling strategies that are able to account for the mechanisms of dislocation/precipitate interaction. In the case of very small precipitates ($<$ 10 nm), the analyses can be based on atomistic simulations, which can establish whether the precipitates are sheared or by-passed by the dislocations, and determine the associated energy barrier \citep{SW10, BTM11, SNL16}. This becomes, however, impossible for larger precipitates ($>$ 100 nm), which are overcome by the formation of dislocation loops, due to computational reasons. The analysis of this process was pioneered by Orowan using a constant line tension model, which computed the critical resolved shear stress (CRSS) necessary to overcome a periodic square array of spherical precipitates impenetrable for dislocations \citep{Orowan1948}. Later, \cite{Bacon1973_BKS} included the effect of the interaction stresses between the dislocation segments, while other authors expanded the results of Orowan to deal with random distributions of  obstacles \citep{Foreman1966_FM, Kocks1966}.

While these approaches can provide {\it qualitative} trends, they cannot be quantitative because the precipitate geometry and orientation as well as the details of the dislocation/precipitate interaction are not taken into account and numerical approaches have been used in recent years. \cite{Xiang2004_LevelSet,Xiang2006_LevelSet} used a level-set representation of the dislocation line to simulate the interaction of both edge and screw dislocations with spherical precipitates. Matrix and precipitate were elastic and isotropic with the same elastic constants and it was assumed that the precipitates could or could not be sheared by dislocations, the latter by including a strong repulsive force on the dislocation within the precipitate. Moreover, the effect of a misfit dilatational strain between the matrix and the precipitate was included. The simulations showed that the richness and complexity of the  dislocation/precipitate interactions and postulated new by-pass mechanisms. However, it should be noted that these simulations did not take into account the crystallography of slip, leading to limitations in the precise modelling of the dislocation mobility (cross-slip, climb).

The influence of crystallographic slip was taken into account by \cite{Monnet2006_DDD_precs} and \cite{Monnet2011_DDD_precs}, who  used discrete dislocation dynamics (DDD) simulations to obtain the CRSS necessary to overcome spherical precipitates, which were made impenetrable to the dislocation by adding a friction stress within the precipitate. Further DDD simulations included the effect of the image stresses induced by the elastic modulus mismatch between the matrix and the spherical precipitate on the CRSS \citep{Takahashi2008_DDD_BEM, Takahashi2011_DDD_FEM, Shin2003_DDD_FEM} using the superposition technique developed by \cite{VanderGiessen1995_DDD_FEM}.  \cite{Takahashi2008_DDD_BEM} reported  the influence of the shear modulus mismatch on the CRSS when the spherical precipitates were sheared  while \cite{Shin2003_DDD_FEM} address  the case of the formation dislocation loops. The influence of the image stresses was higher in the former case but  increased in the latter as several dislocations loops were around the precipitate, increasing the hardening rate. However, the superposition method requires to solve the elastic boundary value problem (using either the finite element or the boundary element method) in each time step of the DDD simulations to obtain the image stresses. Further DDD simulations  to study dislocation/precipitate interactions (many of them focussed in the case of shearable $\gamma'$ precipitates in Ni-based superalloys) have ignored the modulus mismatch \citep{YKN06, VDR09, YZH13, HER13, HZT12, ZSD17, monnet2015multiscale}.  Only more recently, \cite{GFM15} carried out DDD simulations of dislocation/precipitate interaction that took into account the effect of modulus misfit as well as of the misfit stresses arising from the lattice mismatch between $\gamma$ and and $\gamma'$ phases in Ni-based superalloys. The Fast Fourier Transform (FFT) method -- much faster than the traditional methods -- was used in this case to compute the image stresses.

While these results have improved our understanding of precipitation strengthening, they often ignore the actual details of the precipitate shape and orientation, of the dislocation mobility as well as of  the complex stress field around the precipitates (misfit and stress-free transformation strains). More recent analyses  are trying to overcome these limitations by obtaining this critical information from simulations at lower length scales. For instance,  \cite{lehtinen2016multiscale} carried out DDD simulations of dislocation/precipitate interaction in  BCC Fe in which the parameters of the simulation (dislocation mobility, shear modulus,  dislocation core energy) were obtained from atomistic simulations. The precipitates were spherical  and the   dislocation-precipitate interaction was modelled by a Gaussian potential that was calibrated from atomistic simulations. 

This investigation presents a comprehensive multiscale modelling strategy based on DDD to study the mechanisms of dislocation/precipitate interaction. The methodology is applied to Al-Cu alloys but it is general and can be extended to any other metallic alloy. The details of the the $\theta'$ (Al$_2$Cu) precipitates (size, shape and orientation) as well as the stress-free transformation strains around the precipitate were obtained in a previous investigation by the coupling of {\it ab initio} and atomistic simulations with computational thermodynamics and phase-field models \citep{LBL17} and were in good agreement with the experimental data. In addition, the elastic constants and the dislocation mobility laws were obtained from atomistic simulations and this information was used to determine the actual mechanisms of dislocation/precipitate interaction in this system by means of DDD simulations in which all the relevant physical processes were accounted for. In particular, the influence of the precipitate shape, orientation, modulus mismatch and stress-free transformation strain on the dislocation/precipitate interaction mechanisms was analyzed and their influence on the CRSS was determined and compared with the predictions of the classical models for dislocation/precipitate interaction.  

The paper is organized as follows:  the characteristics of the dislocations in the Al matrix and of the $\theta'$ precipitates (obtained using different simulations strategies) are presented in section \ref{sec:Mat_system}, while the DDD simulation strategy is detailed in section \ref{sec:Methodology}. The results of the DDD simulations are shown and discussed in Section \ref{sec:Results}  while the conclusions are drawn in section \ref{sec:Conclusions}.

\section{Material system}
\label{sec:Mat_system}

\begin{figure}[h]
\centering
\includegraphics[scale=0.7]{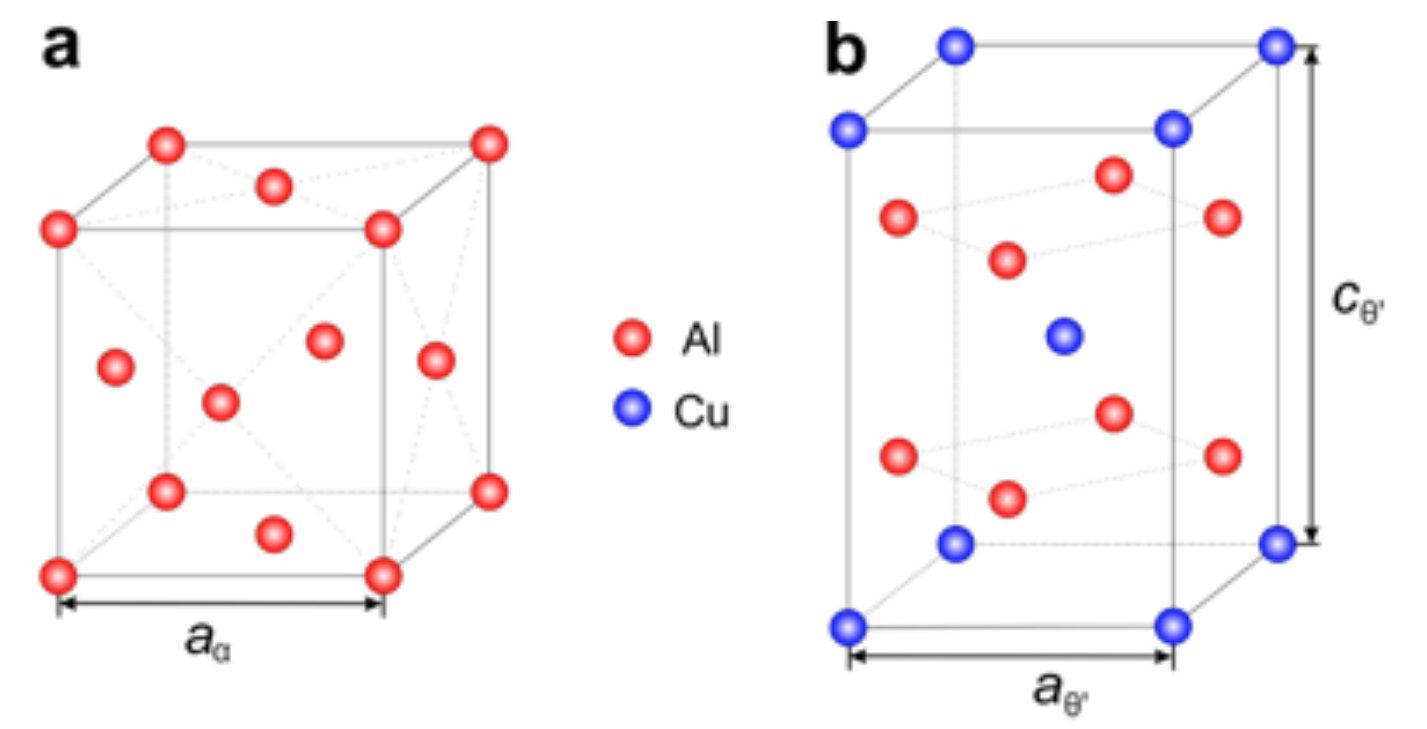}
\caption{Crystal structure of: (a) FCC $\alpha$-Al. (b) BCT $\theta '$ precipitates. \citep{N14}. Red and blue spheres stand for Al and Cu atoms, respectively. From \cite{LBL17}.}  
\label{Crystal}
\end{figure}

\begin{figure}[h]
\centering
\includegraphics[scale=0.5]{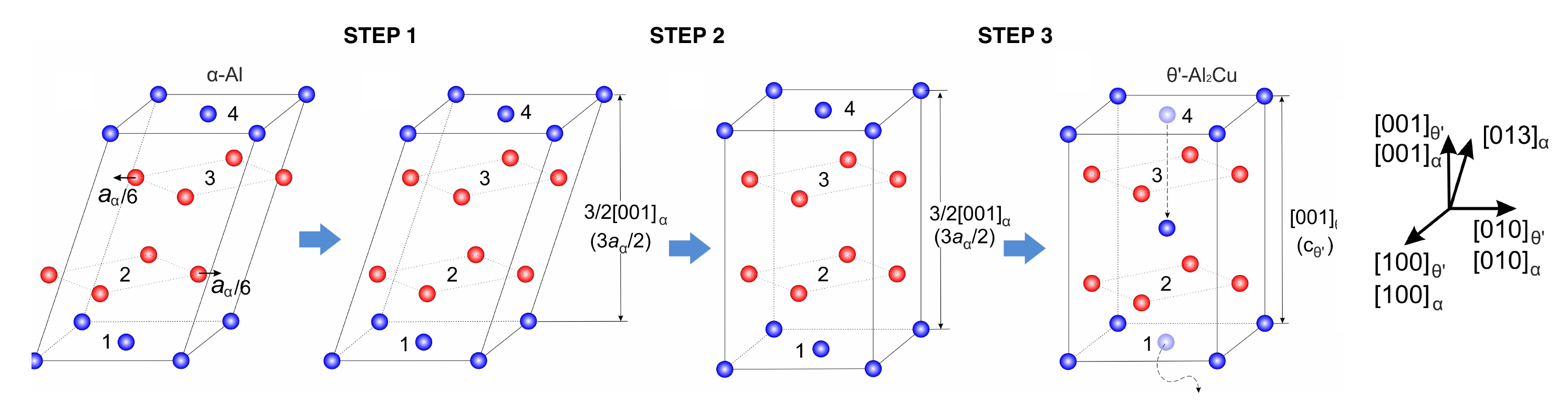}
\caption{Transformation path from $\alpha$-Al to $\theta'$ precipitates during high temperature ageing \citep{N14}. Red and blue spheres stand for Al and Cu atoms, respectively. From \cite{LBL17}.}  
\label{Tpath}
\end{figure}

$\theta'$ precipitates are the key strengthening phase in Al-Cu alloys aged at high temperature (\cite{P95}).  $\theta'$ is a stoichiometric phase with chemical composition Al$_2$Cu and tetragonal structure (space group $I4/mmm$, $a_{\theta'}$ = 0.404 nm, $c_{\theta'}$ = 0.580 nm). The unit cells of $\alpha$-Al (space group $Fm\bar 3m$, $a_\alpha$ = 0.404 nm) matrix and $\theta'$ are shown in Figs. \ref{Crystal} a) and b), respectively. Previous studies of the transformation path of the $\theta'$ precipitate from the $\alpha$-Al lattice have shown three successive steps, which are shown in Fig. \ref{Tpath} \citep{DW83,NM99, N14}. The Al atoms in layers 2 and 3 of the $\alpha$-Al lattice are first shifted in opposite directions by a distance $a_\alpha$/6. This step is followed by a homogeneous shear deformation of the  cell by an angle arctan(1/3) and, finally,  by the shuffling of one Cu atom to the center of the cell and diffusion of the other Cu atom into the Al matrix. According to this transformation path, the lattice correspondence between the parent phase ($\alpha$-Al) and  the $\theta'$ precipitates is given by [013]$_\alpha$ $\rightarrow$ [001]${_\theta'}$ and [010]$_\alpha$ $\rightarrow$ [010]${_\theta'}$ and the transformation matrix, $\mathbf{T}$, that relates the lattice parameters in the $\alpha$-Al, $\mathbf{e}_\alpha$, and in the $\theta'$ precipitate, $\mathbf{e}_{\theta'}$ ($\mathbf{T} \mathbf{e}_\alpha = \mathbf{e}_{\theta'}$)  is expressed as \citep{GLS12, LBL17}

\begin{equation}
\begin{array}{ccc}
\mathbf{T}=\begin{pmatrix}
    a_{\theta'}/a_\alpha &0&0\\  0&a_{\theta'}/a_\alpha&-1/3\\ 0&0&c_{\theta'}/1.5a_\alpha
  \end{pmatrix}.
\end{array}
\label{eq:T}
\end{equation}

The transformation matrix includes both strains and rigid body rotations and the corresponding stress-free transformation strain (SFTS), $\boldsymbol{\epsilon}^0$, can be computed as 

\begin{equation}
\boldsymbol{\epsilon}^0 = \frac{1}{2}({\mathbf T}^T {\mathbf T} - {\mathbf I})
\label{eq:SFTS}
\end{equation}

\noindent where ${\mathbf I}$ stands for the identity matrix.

The nucleation and growth of $\theta'$ precipitates during high temperature ageing has been recently modelled in 3D by means of a multiscale phase-field approach, that takes into account the contribution of the chemical free energy, the interface energy and the elastic energy due to the SFTS \citep{LBL17}.  The chemical free energy was given by computational thermodynamics results, while the interface energy and the lattice parameters of both phases (which determine the elastic energy associated to the SFTS) were obtained from density functional theory simulations. The computed lattice parameters were $a_\alpha$ = 0.405 nm, $a_{\theta'}$ = 0.408 nm, $c_{\theta'}$ = 0.5701 nm, very close to the experimental data reported above \citep{N14} and it was assumed that $a_\alpha \simeq a_{\theta'}$ to compute the SFTS. The multiscale simulation predicted that the $\theta'$ precipitates grew with an orientation relationship (001)$_\alpha$ $\parallel$ (001)$_{\theta'}$, [100]$_\alpha$ $\parallel$ [100]$_{\theta'}$. The precipitates were circular disks and the broad face of the disk was coherent with the Al matrix and  parallel to  either the (100), (010) or (001) planes of the FCC Al lattice, leading to three different orientation variants, while the edges of the circular plates were semi-coherent. Four different deformation variants were possible for each orientation variant of the precipitate due to the four-fold symmetry of the (100) planes in the FCC lattice, leading to a total of 12 deformation variants, in agreement with the experimental observations \citep{DW83,N14}, which were characterized for their corresponding SFTS matrix that can be found in \cite{LBL17}. The simulations predicted a precipitate diameter in the range  120-180 $\mu$m and a thickness  of 4-8$\mu$m, with an average aspect ratio of $\approx$ 26.  These results were in close agreement with experimental data in the literature for peak-aged Al- 4 wt. \% Cu alloys \citep{LBL17, ZCS00, BSW11}.

The elastic constants of the $\alpha$-Al matrix and of the $\theta'$ precipitates were determined using Density Functional Theory (DFT) with the Quantum Espresso plane-wave pseudopotential code \citep{QE_2009}. The exchange-correlation energy was evaluated with the help of the Perdew-Burke-Erzenhof  approach \citep{Ernzerhof1996} within the generalized gradient approximation. Ultrasoft pseudopotentials were used to reduce the basis set of plane wavefunctions used to describe the real electronic functions \citep{Vanderbilt1990}. After careful convergence tests, a cutoff of 37 Ry 
was found to be sufficient to reduce the error in the total energy below 1 meV/atom. A k-point grid separation of 0.03 \AA$^{-1}$ was employed for the integration over the Brillouin zone according to the Monkhorst-Pack scheme \citep{Pack1976}. 

The elastic constants were obtained by applying a given strain to the unit cell in the ground state and calculating the corresponding stress after the atom coordinates in the unit cell were relaxed. Taking into account the crystal symmetries, the cubic $\alpha$- Al unit cell was deformed in the direction normal to the cube face and in shear to compute the three independent elastic constants. The BCT cell of the $\theta'$ precipitate was deformed along two normal directions perpendicular to two faces of the tetragonal lattice and in two shear directions to compute the six independent elastic constants. Six strain levels (varying from -0.003 to 0.003) were used for each deformation pattern to obtain a reliable linear fit of the stress-strain relationship. The elastic constants of $\alpha$-Al and of  $\theta'$ precipitates obtained by DFT  are depicted in Tables \ref{Tab:ECA} and \ref{Tab:ECT}, respectively. The ones for $\alpha$-Al were very close to the experimental data in the literature \citep{VMS64, SS02}. To the best of the authors' knowledge, no experimental data are available for $\theta'$.

\begin{center}
\begin{table}[h]
\begin{center}
\begin{tabular}{cccc}
\hline
  & $C_{11}$ & $C_{12}$ & $C_{44}$\\
\hline
DFT & 110.4  & 60.0 & 31.6 \\
Experimental  & 114.3 & 61.9 & 31.6\\
\hline
\end{tabular}
\end{center}
\caption{Elastic constants (in GPa) of $\alpha$ - Al at 0K obtained from DFT. The experimental values extrapolated at 0K \citep{SS02} are included for comparison.}
\label{Tab:ECA}
\end{table}
\end{center}

\begin{center}
\begin{table}[h]
\begin{center}
\begin{tabular}{cccccc}
\hline
 $C_{11}=C_{22}$ & $C_{12}$ & $C_{13}=C_{23}$ & $C_{33}$ & $C_{44}=C_{55}$ & $C_{66}$\\
\hline
 212.6 & 39.9 & 61.4 & 173.5 & 82.8 & 44.8\\
\hline
\end{tabular}
\end{center}
\caption{Elastic constants (in GPa) of  $\theta'$ - Al$_{2}$Cu precipitates at 0K obtained from DFT. }
\label{Tab:ECT}
\end{table}
\end{center}

\section{Discrete Dislocation Dynamics Strategy}
\label{sec:Methodology}

The dislocation/precipitate interaction is analyzed by means of DDD simulations following the discrete continuous model developed by \cite{Lemarchand2001_DCM}. In this approximation, the dislocations are treated as plate-like inclusions with an eigenstrain that corresponds to the plastic strain associated with the area sheared by the dislocation. The dislocation loops are discretized in segments which move depending on the stresses acting on the segments and the mobility rules and  the plastic strain is computed from the dislocation glide. The DDD code was coupled in the original model with a finite element code that computed the displacement field that is solution of the boundary value problem taking into account the plastic strain provided by the DDD simulations (\cite{Lemarchand2001_DCM}). This framework neither requires the use of analytical expressions for the displacement fields of the dislocation segments (and, thus, can be easily extrapolated to anisotropic materials), nor the computational power increases with the square of the number of dislocations segments. However, computational efforts are limited by the fine finite element discretizations necessary to achieve accurate results, particularly in the case of precipitates with very large aspect ratio. These limitations were overcome recently by \cite{Capolungo2015_DDDFFT}, who used the Fast Fourier Transform (FFT) method to compute the mechanical fields and solve the boundary value problem for periodic cases. Moreover, the  heterogeneous stress distribution that appear due the elastic modulus mismatch between the matrix and the precipitate and the stresses induced by SFTS can be easily incorporated to the simulations.

Dislocations are discretized into segments. The dislocation mobility follows a viscous linear law, where the velocity of node $i$, $\mathbf{v}_i$, of the dislocation line is given by

 \begin{equation}
\mathbf{F}_i = B\mathbf{v}_i
\label{eq:mobility}
\end{equation}

\noindent where $B$ is viscous drag coefficient that depends on the dislocation character and $\mathbf{F}_i$ the force acting on node $i$, which is given by

 \begin{equation}
\mathbf{F}_i = \sum_j \mathbf{f}_{ij}
\end{equation}

\noindent where $\mathbf{f}_{ij}$ is the force acting on the segment $ij$, which is computed according to

 \begin{equation}
\mathbf{f}_{ij} = \int_{\mathbf{x}_i}^{\mathbf{x}_j} N_i(\mathbf{x}) \mathbf{f}_{ij}^{pk}(\mathbf{x}) \rd \mathbf{x}
\end{equation}

\noindent where $N_i$ is the interpolation function associated to node $i$ and $\mathbf{f}_{ij}^{pk}$ is the Peach-Koeler force given by

 \begin{equation}
\mathbf{f}_{ij}^{pk}(\mathbf{x}) =  \Big(\boldsymbol{\sigma}(\mathbf{x}) \cdot  \mathbf{b}_{ij}\Big) \times \mathbf{\hat t}_{ij}
\end{equation}

\noindent where $\mathbf{b}_{ij}$ is the Burgers vector of the segment $ij$ and $\mathbf{\hat t}_{ij}$ the unit vector parallel to the dislocation line. 

\subsection{Calculation of the stress fields}\label{sec:SF}

The stress field  within the simulation domain is computed using the FFT algorithm. The mechanical state of the system is determined by solving the mechanical equilibrium equations  in the domain $V$ with periodic boundary conditions according to

\begin{equation} \label{MechEq}
 \left\{
 \begin{array}{l}
  \tensor{\sigma}(\mathbf{x})=\ftensor{C}(\mathbf{x}):[\tensor{\epsilon}(\mathbf{x})-\tensor{\epsilon}^p(\mathbf{x})-\delta(\mathbf{x})\tensor{\epsilon}^0], \quad \forall \mathbf{x} \in V
\\
 \mathrm{div} ( \tensor{\sigma}(\mathbf{x}))=0  \quad \mathbf{x} \in V \\
\tensor{\sigma}  \cdot \mathbf{n} \text{ has opposite orientation on opposite sides of } \partial V \\
\int_V \tensor{\epsilon}(\mathbf{x})=\mathbf{E}
 \end{array}
 \right.
\end{equation}

\noindent where $\ftensor{C}$ denotes the fourth order elasticity tensor, $\tensor{\epsilon}$ the total strain, $\tensor{\epsilon}^p$ the plastic strain, $\tensor{\epsilon}^0$ the SFTS, $\partial V$ stand for the boundaries of domain $V$ with normal $\mathbf{n}$ and $\mathbf{E}$ is the imposed macroscopic strain. The SFTS is an homogeneous eigenstrain within the precipitate (that only depends on the precipitate variant) and, thus,  $\delta(\mathbf{x})$ is a Dirac delta function that is equal to 1 when $\mathbf{x}$ is within the precipitate and 0 otherwise. This discontinuous strain field may lead to Gibbs fluctuations when using a FFT solver. They were attenuated by the use of discrete gradient operators in Fourier space (in this particular case, the rotational discrete gradient operator proposed by \cite{W15}). The fluctuations in the simulations were not significant, as shown in the stress fields below.

The plastic strain $\tensor{\epsilon}^p(\mathbf{x})$ is computed directly from dislocation motion in the DCM \citep{Capolungo2015_DDDFFT}, while $\tensor{\epsilon}^0$ is given in Section 2. The polarization scheme proposed by \cite{Moulinec1998_FFT} is used to solve the mechanical equilibrium problem in each step of the DDD simulation. This is achieved through the introduction of a reference medium with stiffness $\ftensor{C}^0$ 

\begin{equation}\label{Ref_polar}
  \tensor{\sigma}(\mathbf{x})=\ftensor{C}^0:\tensor{\epsilon}(\mathbf{x})+\tensor{\tau}(\mathbf{x})
\end{equation}

\noindent where $\tensor{\tau}(\mathbf{x})$ is the polarization tensor, which is given by 

\begin{equation}\label{Polarization_with_SFTS}
\tensor{\tau}(\mathbf{x})=\delta\ftensor{C}(\mathbf{x}):\tensor{\epsilon}(\mathbf{x})-\ftensor{C}(\mathbf{x}): \bigg[ \tensor{\epsilon}^p(\mathbf{x}) + \tensor{\epsilon}^{0}(\mathbf{x})\bigg]
\end{equation}

\noindent where $\delta\ftensor{C}(\mathbf{x})=\ftensor{C}(\mathbf{x})-\ftensor{C}^0$. The SFTS tensor $\tensor{\epsilon}^{0}(\mathbf{x})$ takes the value for the corresponding precipitate variant within the precipitate and it is equal to 0 outside of the precipitate \citep{BC18}.  From the expression for the total stress in \eqref{Ref_polar} and the mechanical equilibrium condition \eqref{MechEq}, the mechanical fields can be obtained using the FFT algorithms detailed in \cite{Capolungo2015_DDDFFT, BC18}.

\subsection{Introduction of single dislocation lines}

\begin{figure}[h]
	\centering
	{\includegraphics[width=0.4\textwidth]{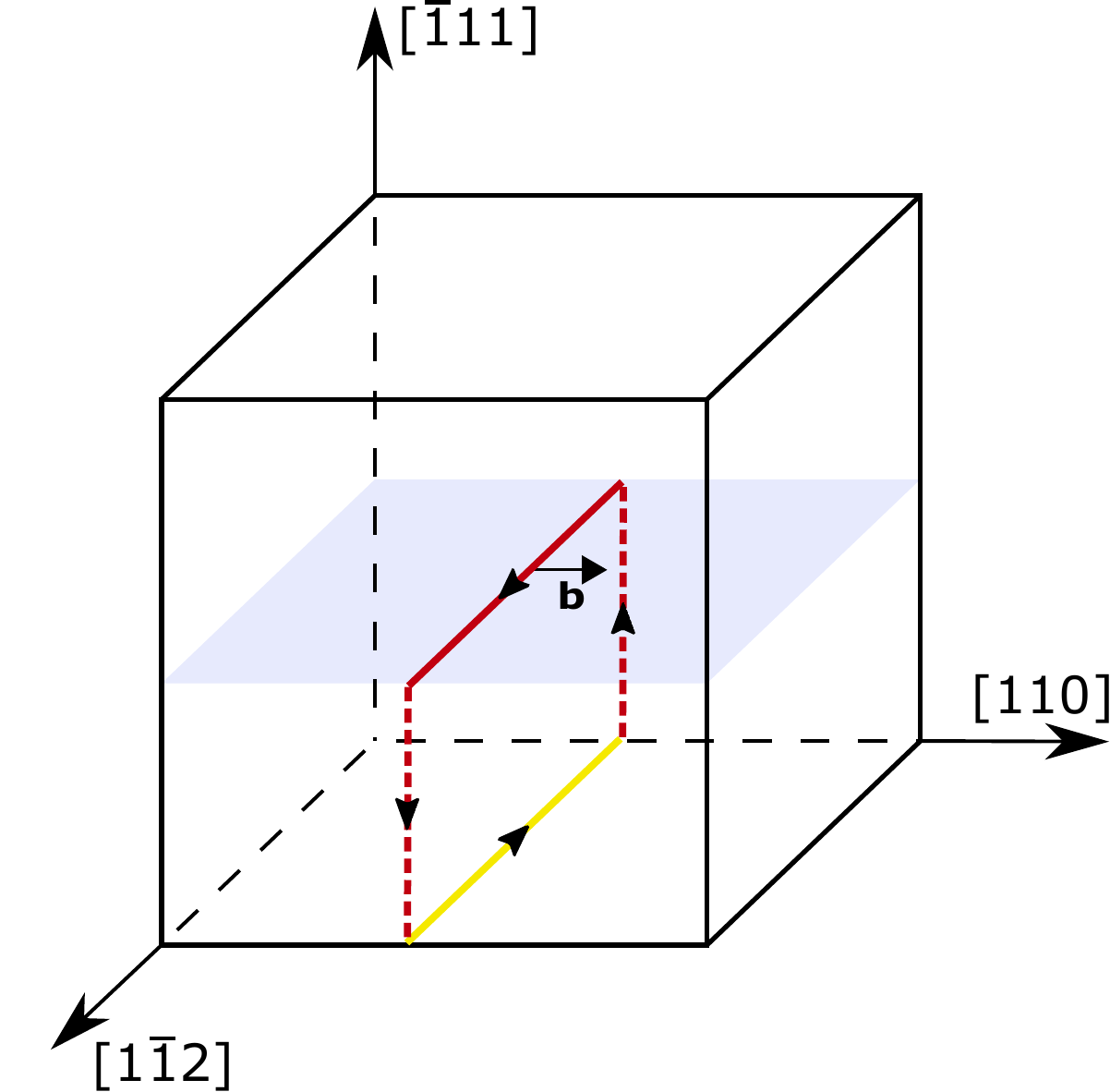}}
\caption{Introduction of a prismatic dislocation loop parallel to one cube faces. Two opposite sides of the prismatic loop (shown as discontinuous lines in the figure) were moved in opposite directions until they reached the boundaries of the domain and annihilate each other, leading to two straight dislocations forming a dipole within the domain. One of the dislocations (yellow line) was fixed during the simulation while the other was free to move in the slip plane (shaded) and interacted with the precipitate.}  
\label{Loop}
\end{figure}

Only dislocation loops can be initially introduced in the DCM  but this  configuration is not appropriate to analyze the interaction of a single dislocation line with the precipitate. This limitation was overcome in the cubic domain (with periodic boundary conditions)  by introducing within the domain a rectangular prismatic loop parallel to one cube faces (Fig. \ref{Loop}). Two opposite sides of the prismatic loop (shown as discontinuous lines in the figure) were moved in opposite directions until they reached the boundaries of the domain and annihilate each other, because they have opposite directions, leading to two straight dislocations forming a dipole within the domain. One of the dislocations was fixed during the simulation (yellow line in the figure) while the other was free to move following the mobility rules established in the following section. The Field Dislocation Mechanics (FDM) method was then used to cancel the stress field created in the domain by the fixed dislocation following the procedure reported in \cite{Berbenni14,Brenner14,Djaka17}.

The FDM method involves the Stokes-Helmholtz decomposition of the elastic distortion into incompatible and compatible parts. The existence of a non-zero dislocation density in the material is accounted for by the incompatible part, while the compatible part ensures that the boundary conditions and the stress equilibrium conditions  are fulfilled \citep{Acharya01}. The incompatible elastic distortion is included in the FDM method through Nye's dislocation tensor $\mathbf{\alpha}$ \citep{Nye53}, which is defined as $\alpha_{ij}=b_it_j$, where $b_i$ is the net Burgers vector in direction $\mathbf{e}_i$ per unit surface $S$ and $t_j$ the dislocation line direction along $\mathbf{e}_j$.

The incompatibility equation and the conservation law are expressed, respectively, by

\begin{equation}
\rm{curl} (\mathbf{U}^e)=\boldsymbol{\alpha}
\end{equation}

\begin{equation}
\rm{div} (\boldsymbol{\alpha})=0
\end{equation}

\noindent where $\mathbf{U}^e$ is the elastic distortion tensor. Applying the Stokes-Helmholtz decomposition to the elastic distortion tensor,

\begin{equation}
\mathbf{U}^e=\mathbf{U}^{e,\bot}+\mathbf{U}^{e,\parallel}
\end{equation}

\noindent where $\mathbf{U}^{e,\bot}$ and $\mathbf{U}^{e,\parallel}$ stand for the incompatible and compatible parts of the elastic distorion respectively. Taking into account that

\begin{equation}
\boldsymbol{\alpha}=\rm{curl} (\mathbf{U}^{e,\bot})
\label{alpha1}
\end{equation}

\begin{equation}
\rm{div} (\mathbf{U}^{e,\bot})=0
\end{equation}

\noindent and applying again the operator curl to the expression \eqref{alpha1}, after some manipulation, it yields a Poisson-type equation

\begin{equation}
\rm{div}(\rm{grad}(\mathbf{U}^{e,\bot}))=-\rm{curl} (\boldsymbol{\alpha})
\label{Poisson1}
\end{equation}

\noindent that can be expressed in component form as
\begin{equation}
U^{e,\bot}_{ij,kk}=-e_{jkl}\alpha_{il,k}
\label{Poisson_FDM}
\end{equation}

The Poisson equation \eqref{Poisson_FDM} can be solved in the Fourier space. To this end, the dislocation density $\mathbf{\alpha}(\mathbf{x})$ is computed in the Fourier space, $\tilde{\mathbf{\alpha}}(\boldsymbol{\xi})$, and the incompatible elastic distortion is obtained in the Fourier space as:

\begin{equation}
\begin{aligned}
\tilde{U}^{e,\bot}_{ij}(\boldsymbol{\xi})&=\mathrm{i}\frac{\xi_k}{\xi^2}e_{jkl}\tilde{\alpha}_{il}(\boldsymbol{\xi}) \quad \forall\mathbf{\xi}\neq \mathbf{0} \\
\tilde{U}^{e,\bot}_{ij}(\mathbf{0})&=\mathbf{0}.
\end{aligned}
\end{equation}

Once $\tilde{\mathbf{U}}^{e,\bot}(\boldsymbol{\xi})$ is known, the inverse Fourier transform is computed to get the incompatible elastic distortion in the real space. Finally, the incompatible elastic strain $\boldsymbol{\epsilon}^{e,\bot}$ is the symmetric part of $\mathbf{U}^{e,\bot}$. The incompatible elastic strain is introduced as plastic strain in the Lippmann-Schwinger equation, which is solved using the FFT algorithm following the same procedure used in section \ref{sec:SF}.

In order to screen the stress field of a straight edge dislocation  with Burgers vector $\mathbf{b}$, the corresponding Nye tensor is given by

\begin{equation}
\begin{array}{ccc}
\boldsymbol{\alpha}=\begin{pmatrix}
   0 &0&0\\  -b/l^2&0&0\\ 0&0&0
  \end{pmatrix} 
\end{array}
\label{eq:alpha}
\end{equation}

\noindent where $-b$ is the magnitude of the Burgers vector (opposite to the one of the dislocation to cancel the stress field) and $l$ the voxel size of the discretization. This value of the Nye tensor is applied in the voxels where the immobile dislocation is located, being zero in the rest of the simulation domain.

\subsection{Dislocation mobility}

\begin{figure}[h]
  \begin{center}
    \includegraphics[scale=0.8]{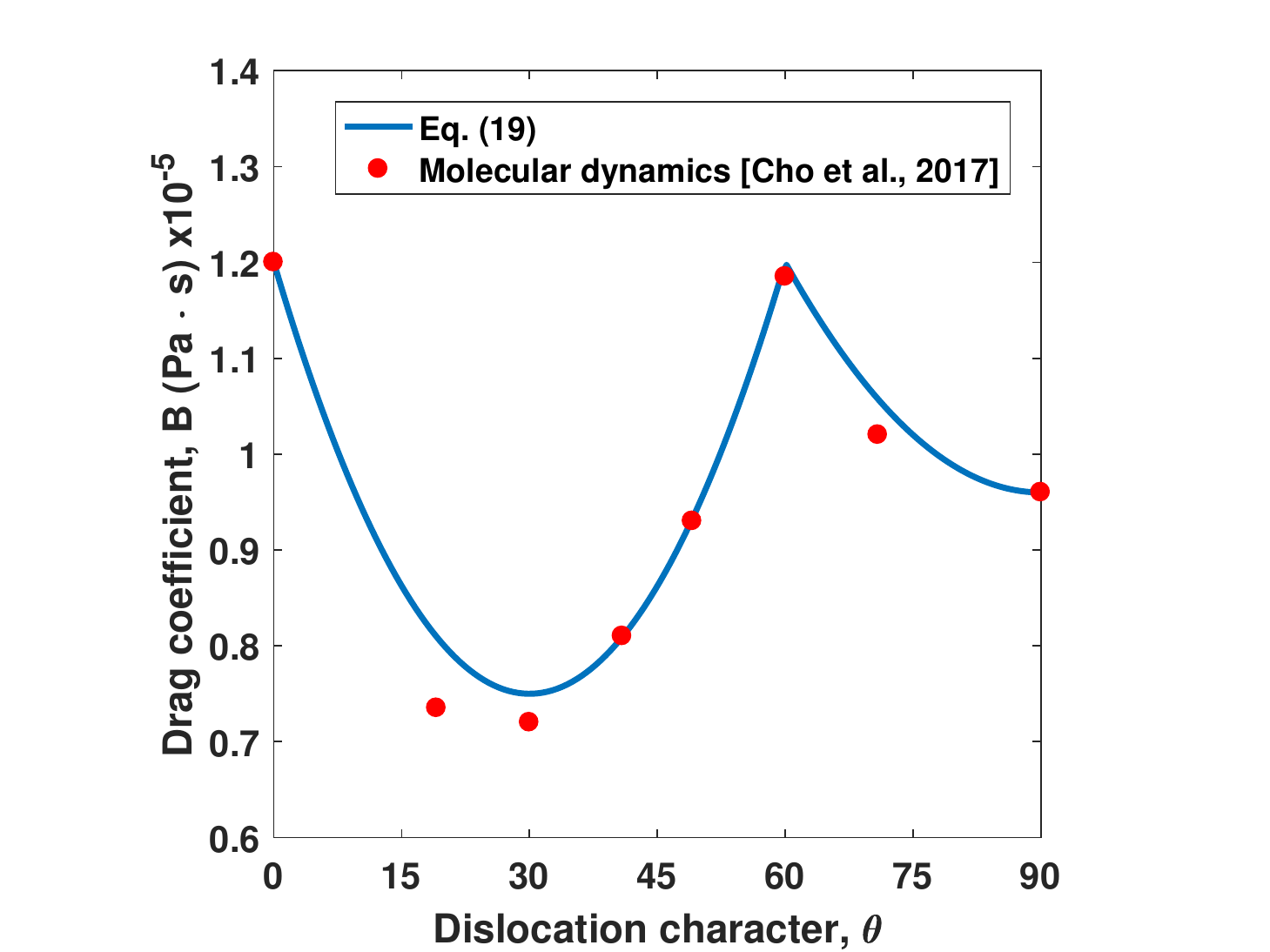}
  \end{center}
  \caption{Drag coefficient $B$ for the dislocation mobility in Al at 300K as a function of the dislocation character.  $\theta$ = 0$^\circ$ stands for pure screw dislocation and $\theta$ = 90$^\circ$ for pure edge dislocation. The solid symbols stand for moecular dynamics simulation in \cite{Molinari2017_Mobility}, while the solid line corresponds to eq. (\ref{eq:DragMatrixCho}).}
\label{fig:MobLawCho}
\end{figure}

The drag coefficient vector, $B$, that characterizes the dislocation mobility  has been recently determined in Al by \cite{Molinari2017_Mobility}. They carried out molecular dynamics simulations of straight dislocation segments with different character and determined ${B}$ as a function of temperature in the regime in which the dislocation mobility is controlled by the viscous friction force arising from phonon damping.  it was found that the drag coefficient of a mixed dislocation cannot be obtained by a linear interpolation between those of edge and screw dislocations (Fig. \ref{fig:MobLawCho}). The maximum drag coefficients were found for the screw dislocation and a  mixed dislocation whose Burgers vector forms an angle of 60$^\circ$ with the dislocation line. The drag coefficients obtained from the molecular dynamics simulations were fitted to two parabolic functions according to

\begin{multline}\label{eq:DragMatrixCho}
B(\theta) = B(0)-\bigg(B(0)-\frac{B(0)}{1.6}\bigg)\frac{12}{\pi}|\theta| +\left(B(0)-\frac{B(0)}{1.6}\right)\frac{36}{\pi^2}\theta^2 \qquad 0<\theta<\frac{\pi}{3} \\
B(\theta) = B(0)+\bigg(B(\pi/2)-B(0)\bigg)\frac{12}{\pi}\left(|\theta|-\frac{\pi}{3}\right)-  \\ -\bigg(B(\pi/2)-B(0)\bigg)\frac{36}{\pi^2}\left(|\theta|-\frac{\pi}{3}\right)^2 \quad \frac{\pi}{3}<\theta<\frac{\pi}{2}
\end{multline}

\noindent where $B(0)$ and $B(\pi/2)$ stand for the drag coefficients of pure screw and edge dislocations, respectively. This drag coefficient was used in the DDD simulations for the Al matrix. The drag coefficient in the $\theta'$ precipitates was assumed to be infinity and, therefore, the dislocations could not shear the precipitates and were forced to by-pass them.

\section{Results and discussion}
\label{sec:Results}

\begin{figure}[h]
\centering
	{\includegraphics[width=\textwidth]{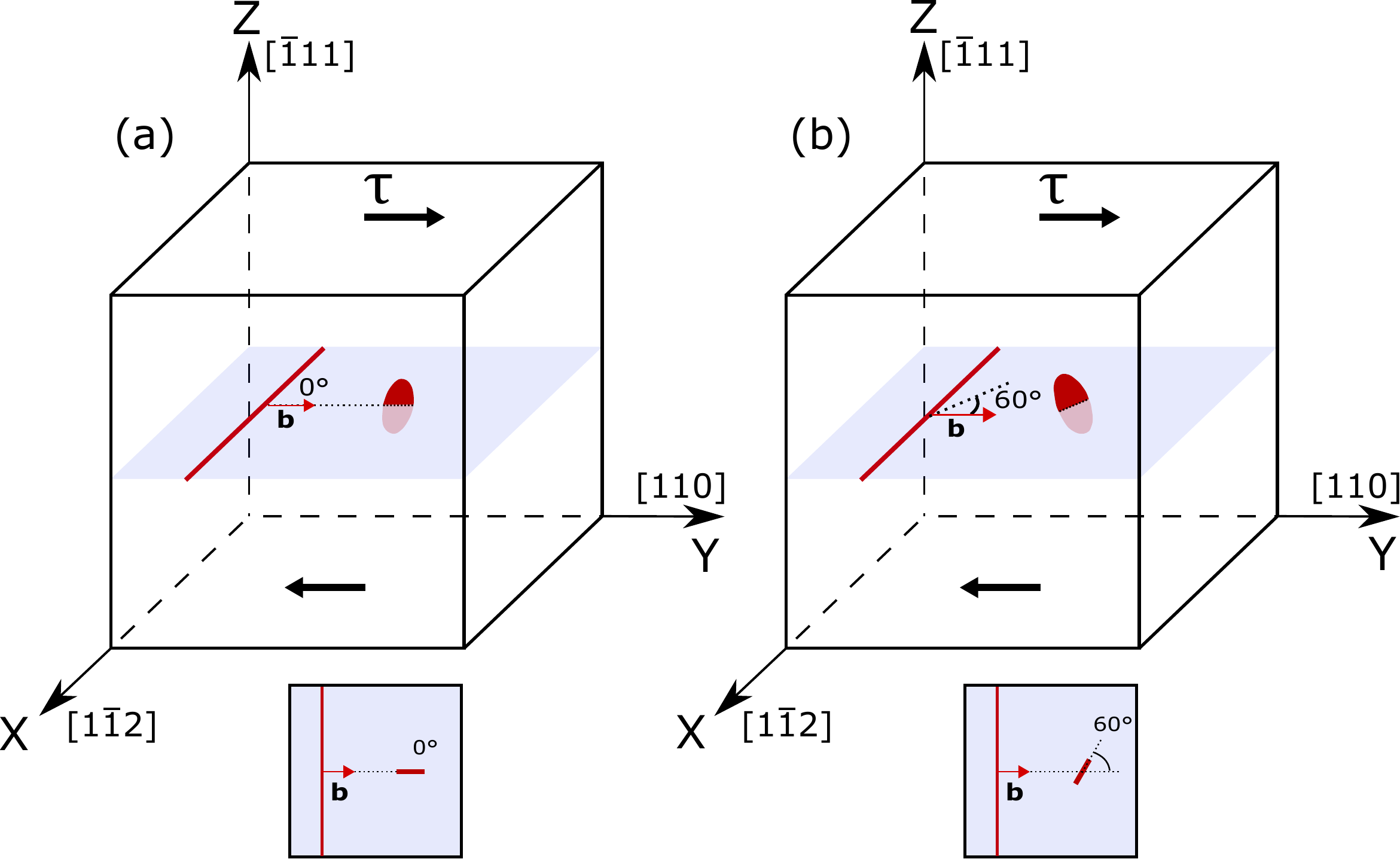}}
\caption{Initial configuration of the edge dislocation and the $\theta'$ precipitate for the DDD simulations. (a) Precipitate habit plane (001). The angle between the Burgers vector and the section of the precipitate along the glide plane is 0$\degree$. (b) Precipitate habit plane (010). The angle between the Burgers vector and the section of the precipitate along the glide plane is 60\degree. The orientation of the dislocation line and the precipitate in the slip plane are shown  for both configurations below each figure.}  
\label{Initial}
\end{figure}

DDD simulations were carried out using a cubic domain of 729 x 729 x 729 nm with periodic boundary conditions, which was discretized with a grid of 128 x 128 x 128 voxels. The axes of the cubic domain were aligned with the [1$\bar1$2], [110] and [$\bar1$11] directions of the Al lattice.  A straight edge dislocation ($\bar1$11)[110] was introduced in the simulation box following the strategy described above. The precipitate was inserted at the center of the simulation box as a circular disk parallel to either the (001) and (010) plane, which stand for the respective habit planes. $\theta'$ precipitates also grow along the (001) habit plane but this dislocation/precipitate configuration is equivalent to the that of the (010) precipitates.   The slip plane of the dislocation intersects the center of the precipitate. The initial configuration is represented in figure \ref{Initial} for both orientations of the precipitate. For the precipitate parallel to the (001) plane, the section of the precipitate along the glide plane was parallel  to the Burgers vector (Fig. \ref{Initial}a), whereas it formed an angle of 60\degree for the precipitate in the (010) plane (Fig. \ref{Initial}b). A shear strain rate is applied to the cubic domain along the [110] direction, as shown in Figure \ref{Initial}.

The precipitate volume fraction was held constant and equal to 3.1 10$^{-4}$ in the simulations, so the interaction between precipitates can be neglected.  The elastic constants of the Al matrix and of the $\theta'$ precipitate in Tables \ref{Tab:ECA} and \ref{Tab:ECT}, respectively, were used, while the dislocation mobility in Al was given by the drag coefficient $B$ in eq. (\ref{eq:DragMatrixCho}). It was assumed that the precipitate was impenetrable to dislocations. All the simulations presented below were carried out at an applied strain rate of 10$^{4}$ s$^{-1}$ because the results obtained at this strain rate are equivalent to those obtained under quasi-static conditions, as shown in the Appendix.

\subsection{Mechanisms of dislocation - precipitate interaction}

The mechanisms of dislocation precipitate interaction and the particular role played by the SFTS around the precipitate can be understood from the shear stress-strain curves obtained from the DDD simulations. In this section, the stress-strain curves and the path followed by the dislocation is analyzed for each precipitate variant. The precipitate diameter in these simulations was 156 nm and the aspect ratio 26:1, in agreement with the results of the phase-field simulations for $\theta`$ in Al-Cu alloys \citep{LBL17}. Simulations were carried with and without
including the effect of the SFTS to assess the influence of this factor on the mechanics of dislocation/precipitate interactions. The interaction of the dislocation with the 12 deformation variants induced by the presence of the SFTS is reduced to 6 independent cases due to the symmetries of the FCC lattice, two corresponding to the 0\degree configuration (Fig. \ref{Initial}a) and four to the 60\degree configuration (Fig. \ref{Initial}b).

\subsubsection{0\degree orientation}

\begin{figure}
\centering
\includegraphics[width=0.6\textwidth]{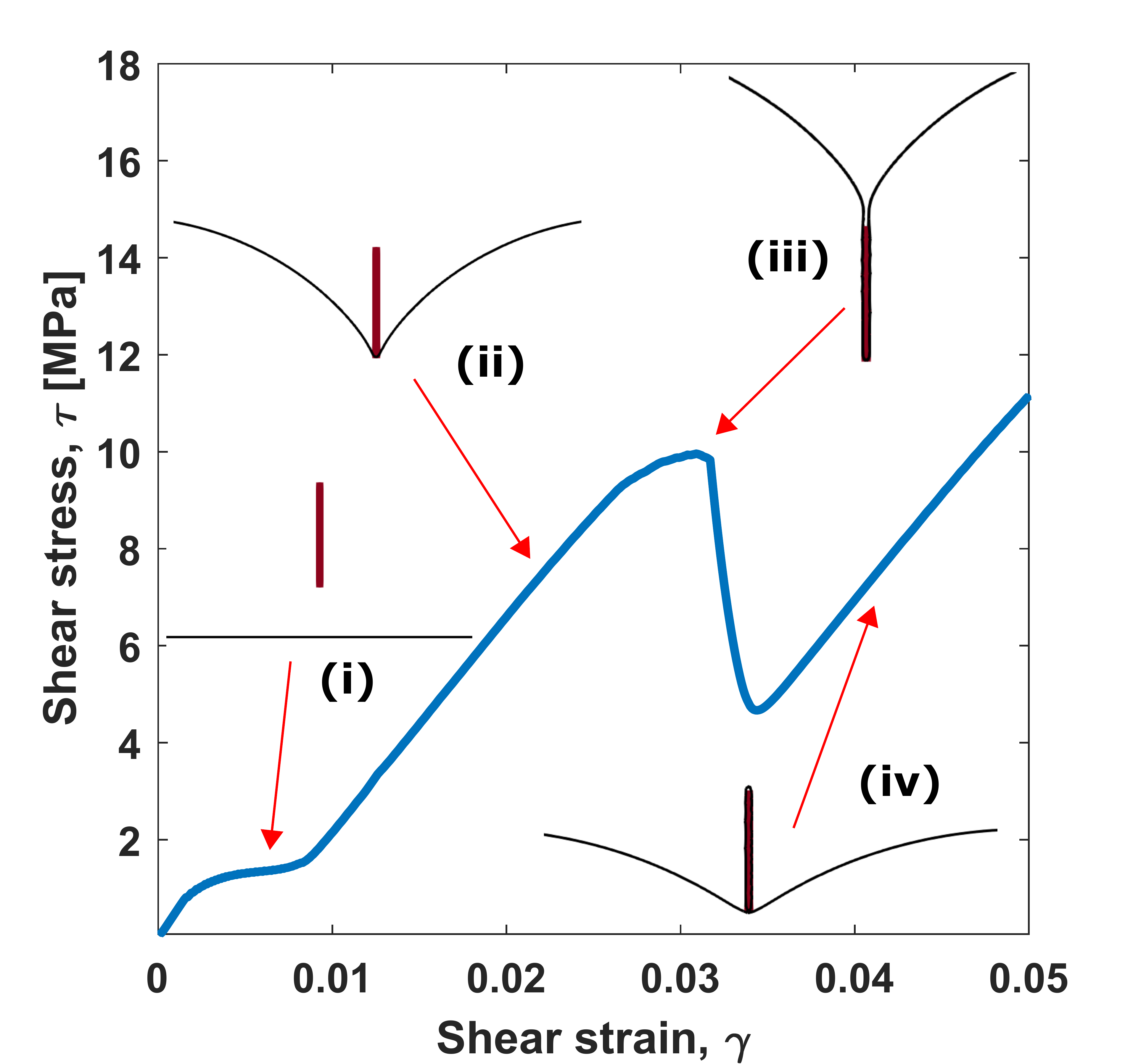}
\caption{Shear stress-strain curve corresponding to the dislocation/precipitate interaction in the 0\degree configuration without SFTS. The evolution of the dislocation line around the precipitate is showed in several points along the curve.}  
\label{V1_NoSFTS}
\end{figure}

The shear stress-strain curve of the simulation in the 0\degree orientation without SFTS is plotted in Fig. \ref{V1_NoSFTS}. The configuration of the dislocation line around the cross-section of the precipitate in the glide plane is also included in the figure for different values of the applied strain. In the initial stages of deformation, marked with (i) in the figure, dislocation glide takes place at very low stress and the dislocation line remains straight, indicating that there is no influence of the precipitate. Linear hardening is observed afterwards in region (ii) as the dislocation starts to bow around the precipitate. The dislocation overcomes the precipitate by the formation of an Orowan loop, as shown in (iii) and, as the dislocation leaves the domain,  another dislocation enters the domain by the opposite boundary due to the periodic boundary conditions (region iv), leading to hardening.

The dislocation precipitate interaction for the 0\degree configuration is changed in the presence of the stress fields around the precipitate induced by the SFTS, which can be found in Table 1 of \cite{LBL17}. The magnitude of the SFTS in this table is given in a reference system which follows the orientation relationship between the matrix and the precipitate. They have to be rotated to the reference frame in Fig. \ref{Initial}  and the two SFTS considered for this precipitate configuration are $\tensor\epsilon^{0}_1$ and $\tensor\epsilon^{0}_2$, which are given by \citep{LBL17}. 

\begin{equation}
 \tensor\epsilon^{0}_1= \mathbf{R}^T\begin{pmatrix}
    0&0&0\\  0&0&-0.1667\\ 0&-0.1667&-0.0042
  \end{pmatrix}\mathbf{R} \quad {\rm and} \quad
   \tensor\epsilon^{0}_2=\mathbf{R}^T\begin{pmatrix}
    0&0&0\\  0&0&0.1667\\ 0&0.1667&-0.0042
  \end{pmatrix}\mathbf{R}
\label{SFTS_0}
\end{equation}

\noindent where $\mathbf{R}$ is the rotation matrix, which is expressed as

\begin{equation}
   \mathbf{R}=\begin{pmatrix}
    0.408&-0.408&0.816\\  0.707&0.707&0\\ -0.577&0.577&0.577
  \end{pmatrix}.
\label{Rot_matrix}
\end{equation}

\begin{figure}
\centering
\includegraphics[width=0.49\textwidth]{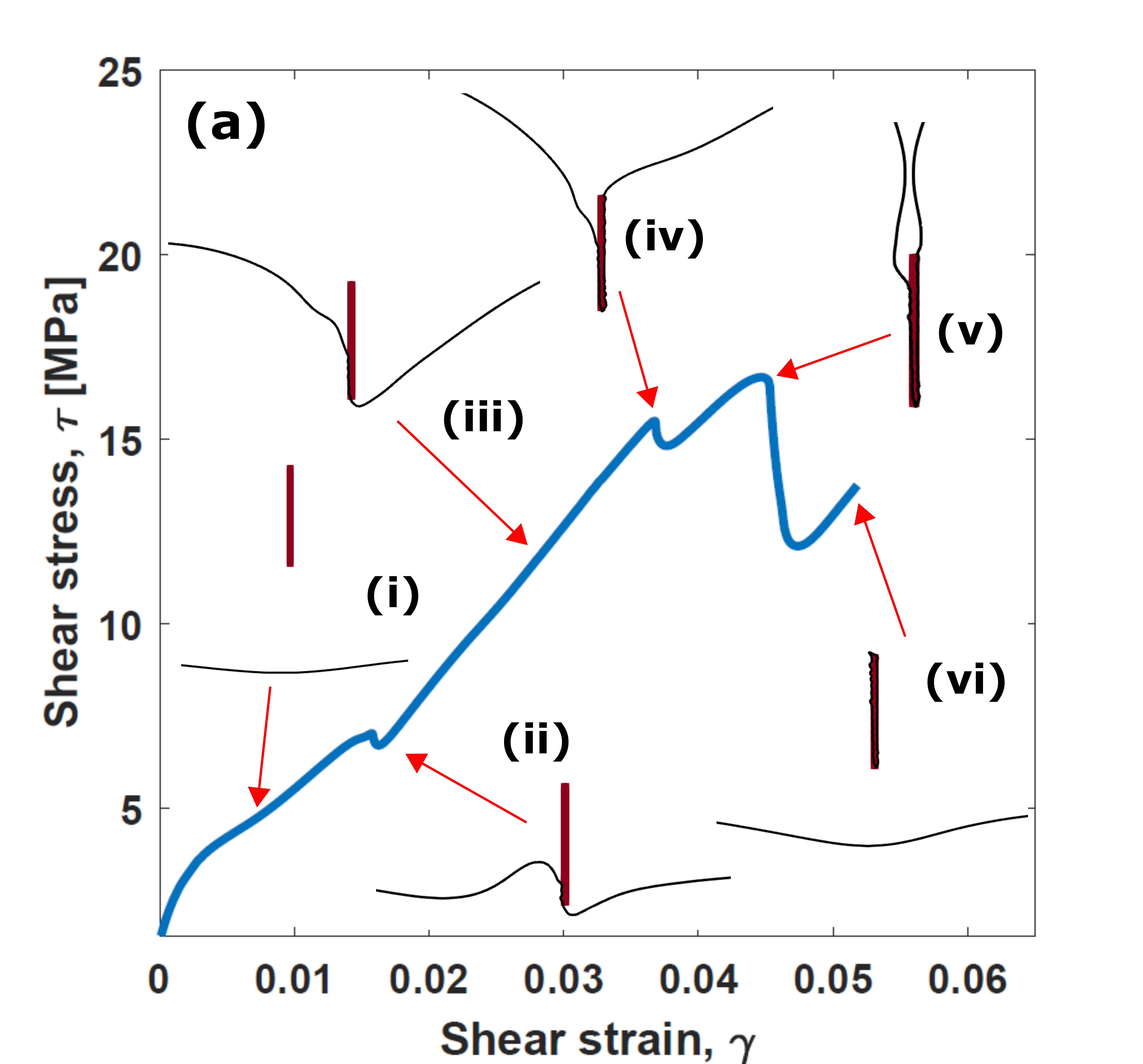}
\includegraphics[width=0.49\textwidth]{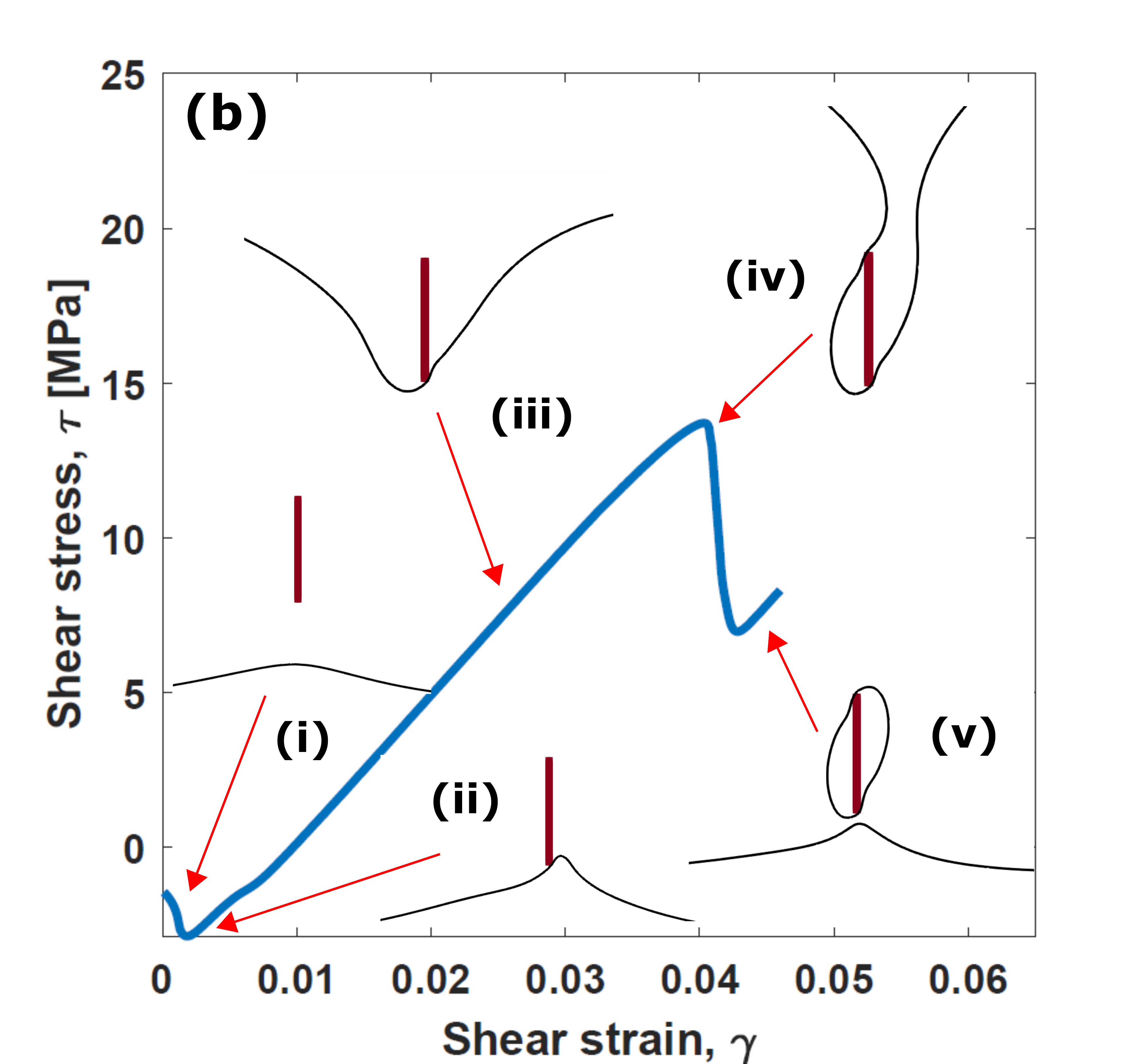}
\caption{Shear stress-strain curve corresponding to the dislocation/precipitate interaction in the 0\degree configuration with SFTS. (a) $\tensor\epsilon^{0}$ = $\tensor\epsilon^{0}_1$. (b) $\tensor\epsilon^{0}$ = $\tensor\epsilon^{0}_2$.The evolution of the dislocation line around the precipitate is showed in several points along the curve.}  
\label{V1_SFTS}
\end{figure}

In the case of $\tensor\epsilon^{0}_1$,  (Fig. \ref{V1_SFTS}a), the stress field around the precipitate leads to an initial repulsion between the dislocation and the precipitate, which is shown in the initial hardening in the stress-strain curve in region (i) and in the shape of the dislocation line. After this initial barrier is overcome, one small segment of the dislocation line is attracted to the precipitate (region ii) and the dislocation starts to bow around the precipitate (region iii) but the dislocation loop is not symmetric due to the SFTS. The Orowan loop is finally created around the precipitate (regions iv and v) and the process is repeated as a new dislocation enters the domain (region vi). The stress field created by the SFTS $\tensor\epsilon^{0}_2$ leads to a different behavior, as shown in Fig. \ref{V1_SFTS}b). The dislocation line is initially attracted to the precipitate (region i) and a minimum in the stress-strain curve is found when the dislocation line gets in contact with the precipitate (region ii). Linear hardening is observed afterwards as the dislocation bows around the precipitate (region iii) and overcomes the obstacle by the formation of an Orowan loop (region iv). However, the final Orowan loop is not attached to the precipitate surface and the final shape of the Orowan loop is different from the ones found in Figs. \ref{V1_NoSFTS} and \ref{V1_SFTS}a).

The presence of the SFTS increased considerably the CRSS (i.e. the maximum stress in the shear stress-strain curve) necessary to overcome the precipitate. According to the line tension model, the CRSS is controlled by the minimum radius of curvature of the dislocation line during the Orowan process, which decreased in the presence of the SFTS because of the anisotropy introduced in the development of the Orowan loops. In addition, the CRSS in the presence of $\tensor\epsilon^{0}_1$ was slightly higher than the one in the presence of $\tensor\epsilon^{0}_2$.

\subsubsection{60\degree orientation}

\begin{figure}[t]
\centering
\includegraphics[width=0.6\textwidth]{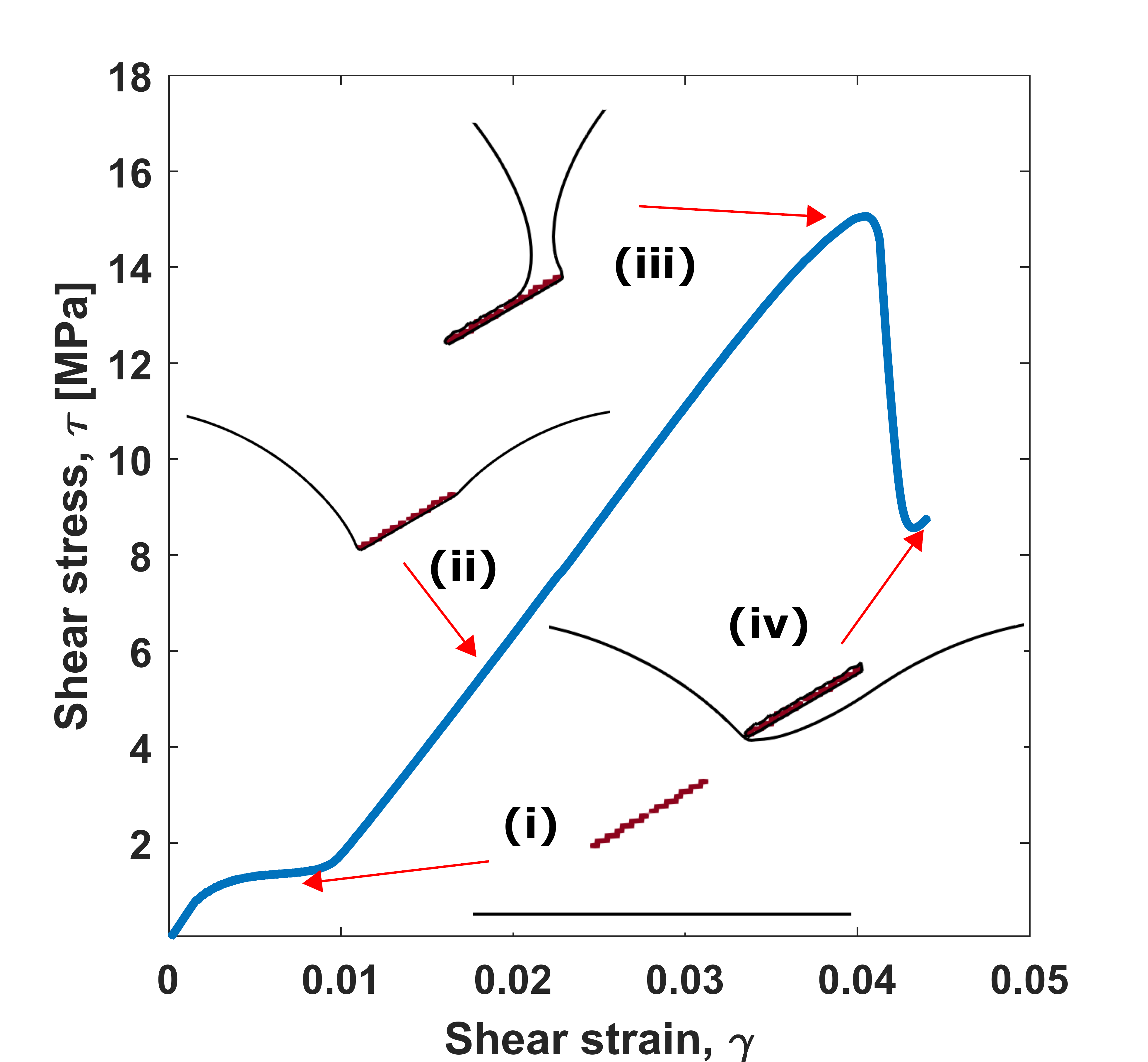}
\caption{Shear stress-strain curve corresponding to the dislocation/precipitate interaction in the 60\degree configuration without SFTS.The evolution of the dislocation line around the precipitate is showed in several points along the curve.}  
\label{V5_NoSFTS}
\end{figure}

Similar DDD simulations were carried out when the precipitate was in 60\degree configuration. In the absence of the SFTS, the dislocation overcomes the precipitate by the formation of an Orowan loop (Fig. \ref{V5_NoSFTS}) and the regions of the shear stress-strain curve are equivalent to those found in Fig.  \ref{V1_NoSFTS} in the 0\degree orientation in the absence of the SFTS. In this case, the dislocation line advances toward the precipitate and rotates until is in full contact with the broad face of the precipitate (region ii). Afterwards, the dislocation arms  advance until an Orowan loop is formed around the precipitate (region iii).

\begin{figure}[!]
\centering
\includegraphics[width=0.49\textwidth]{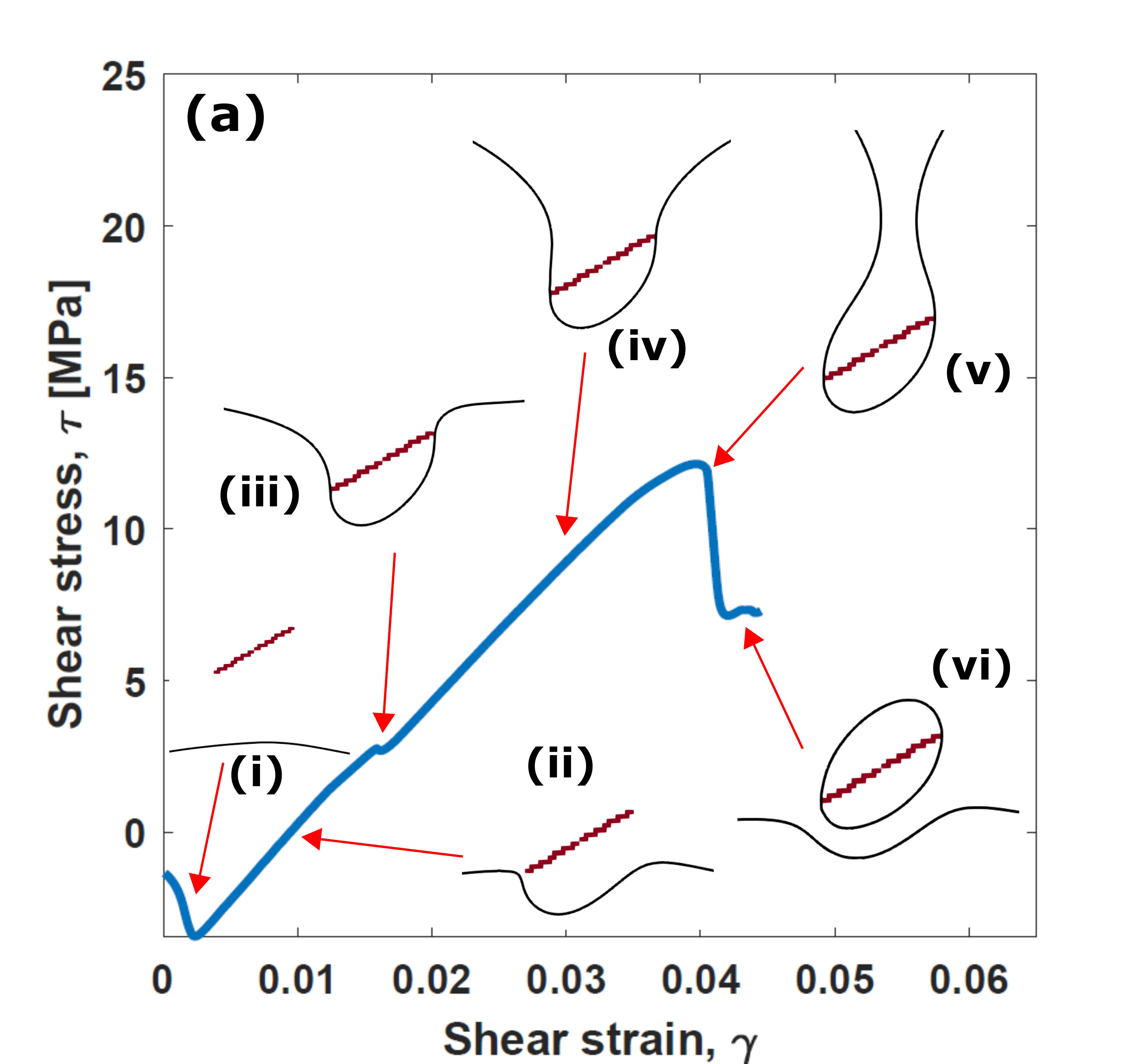}
\includegraphics[width=0.49\textwidth]{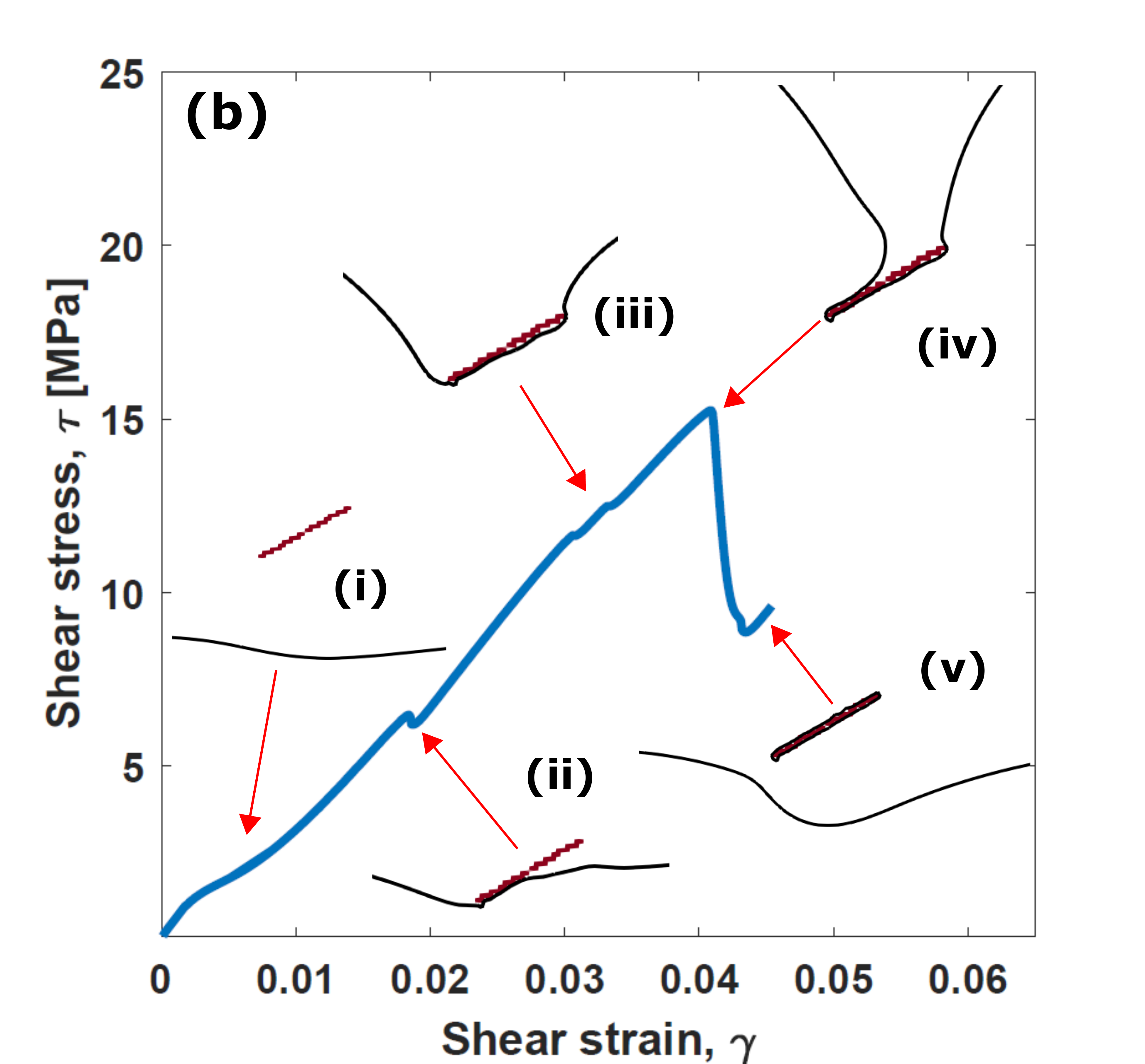}
\includegraphics[width=0.49\textwidth]{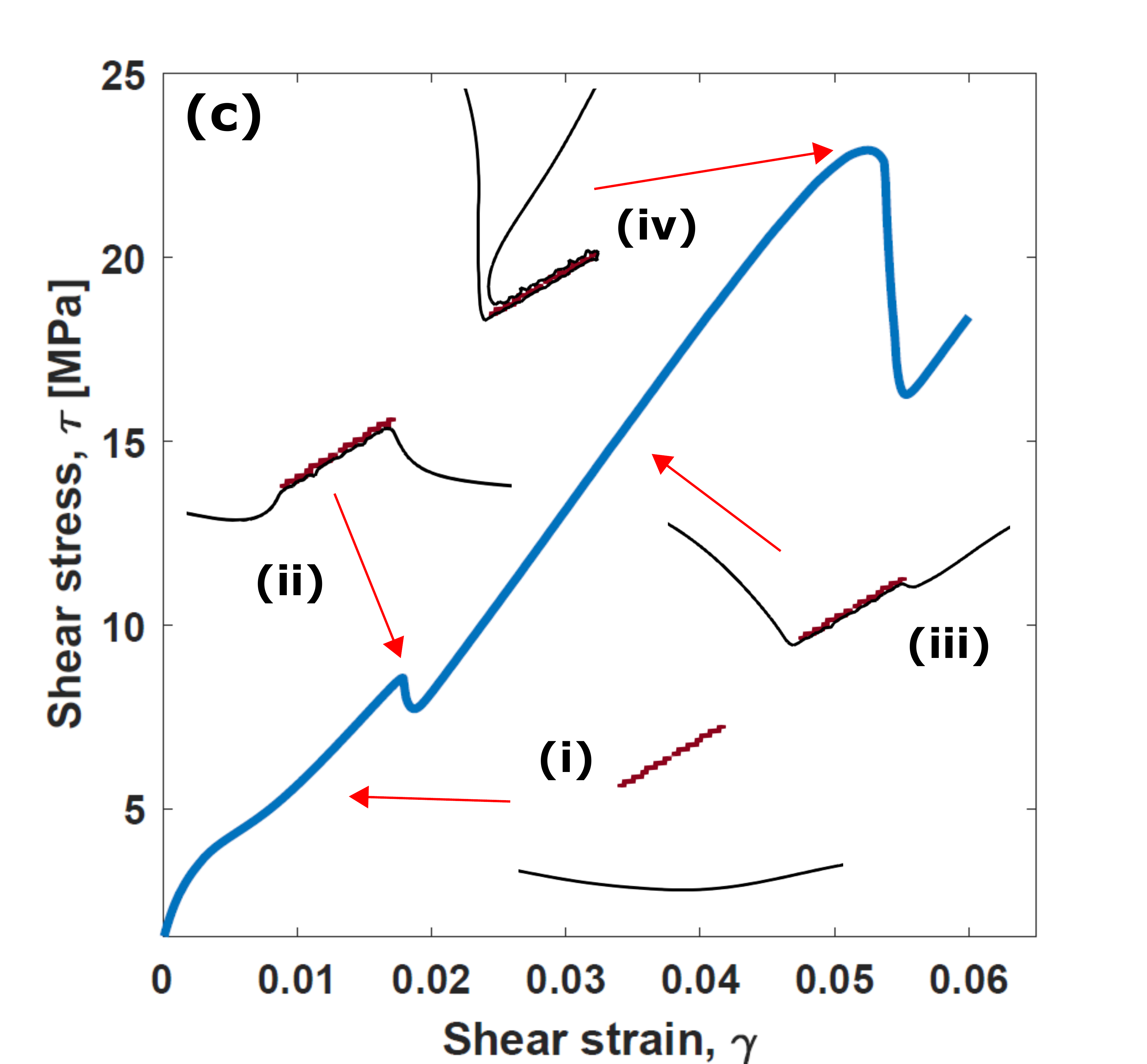}
\includegraphics[width=0.49\textwidth]{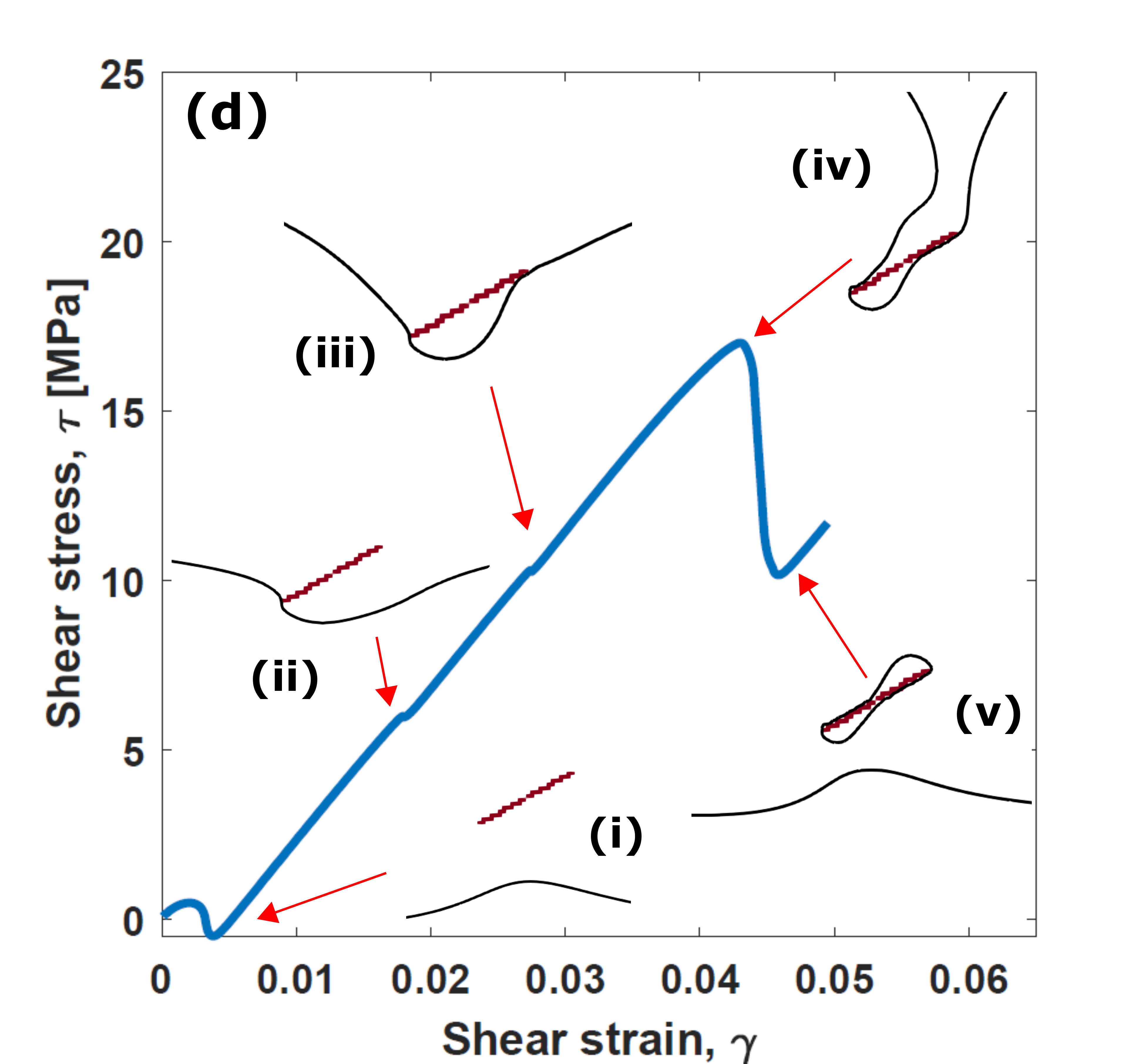}
\caption{Shear stress-strain curve corresponding to the dislocation/precipitate interaction in the 60\degree configuration with SFTS. (a) $\tensor\epsilon^{0}$ = $\tensor\epsilon^{0}_3$. (b) $\tensor\epsilon^{0}$ = $\tensor\epsilon^{0}_4$. (c) $\tensor\epsilon^{0}$ = $\tensor\epsilon^{0}_5$. (d) $\tensor\epsilon^{0}$ = $\tensor\epsilon^{0}_6$. The evolution of the dislocation line around the precipitate is showed in several points along the curve.}  
\label{V5_SFTS}
\end{figure}

In the orientation 60\degree, there are four independent SFTS that lead to changes in the dislocation-precipitate interaction mechanisms. The corresponding STFS are given by

\begin{equation}
\tensor \epsilon^{0}_3=\mathbf{R}^T\begin{pmatrix}
    0&0&0\\  0&-0.0042&0.1667\\ 0&0.1667&0
  \end{pmatrix} \mathbf{R}
\quad {\rm and} \quad
 \tensor \epsilon^{0}_4=\mathbf{R}^T\begin{pmatrix}
    0&0.1667&0\\  0.1667&-0.0042&0\\ 0&0&0
  \end{pmatrix} \mathbf{R}
\label{SFTS_60}
\end{equation}

\begin{equation}
 \tensor\epsilon^{0}_5=\mathbf{R}^T\begin{pmatrix}
    0&0&0\\  0&-0.0042&-0.1667\\ 0&-0.1667&0
  \end{pmatrix}\mathbf{R} \quad {\rm and} \quad  \tensor\epsilon^{0}_6=\mathbf{R}^T\begin{pmatrix}
    0&-0.1667&0\\  -0.1667&-0.0042&0\\  0&0&0
  \end{pmatrix} \mathbf{R}
\label{SFTS_601}
\end{equation}

\noindent and the dislocation-precipitate interactions in the presence of the stress fields created by the SFTS are plotted in Fig. \ref{V5_SFTS}, together with the corresponding shear stress-strain curves. In all cases, the dislocation line tends to rotate and to become parallel to the broad face of the precipitate, and an Orowan loop  is formed afterwards as the dislocation arms propagate at both sides of the precipitate. However, the approximation of the dislocation to the precipitate and the formation of the Orowan loop is modulated by the SFTS.

\begin{figure}[!]
\centering
\includegraphics[width=\textwidth]{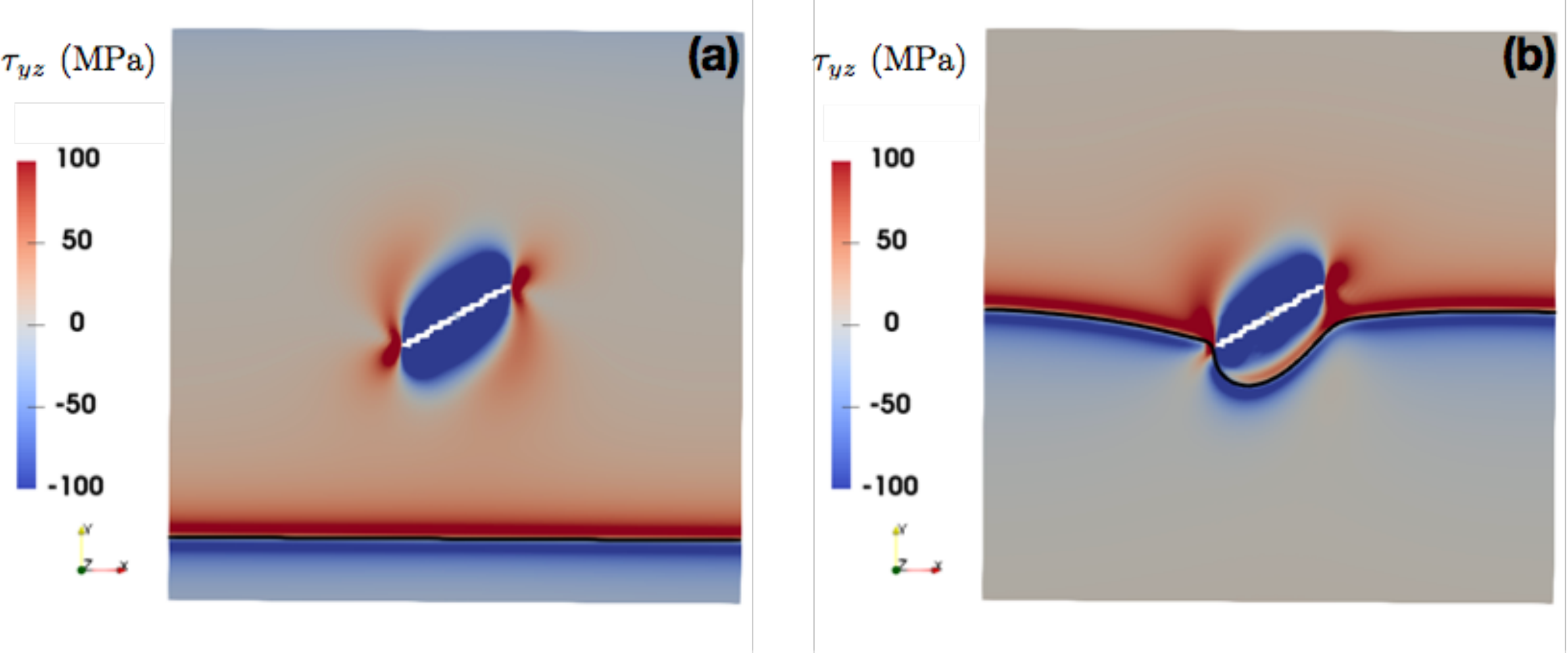}
\caption{Contour plots of the shear stress, $\tau_{yz}$ corresponding to the dislocation/precipitate interaction in the 60\degree configuration with SFTS $\tensor\epsilon^{0}$ = $\tensor\epsilon^{0}_3$.  (a) Initial configuration.  (b) Dislocation/precipitate configuration corresponding to (iii) in Fig. \ref{V5_SFTS}a. The dislocation is shown as a black line and the cross-section of the precipitate is white.}  
\label{Stress-AR26}
\end{figure}

In the case of $\tensor\epsilon^{0}_3$ (Fig. \ref{V5_SFTS}a), the stress field near the precipitate initially attracts the dislocation toward the precipitate (region i), but this is followed by a strong repulsion between the dislocation line and the broad face of the precipitate (region ii),  leading to the formation of a half loop whose extremes are in contact with precipitate (regions iii and iv). The interaction between the stress field of the dislocation and the stress field created by the STFS  in this case is shown in the contour plots of $\tau_{yz}$ in Fig. \ref{Stress-AR26}. The repulsion between the dislocation and the precipitate due to the presence of the STFS leads to the formation of the ellipsoidal Orowan loop which is only in contact with the edges of the precipitate (regions v and vi). Interestingly, the minimum radius of curvature of the dislocation line during the Orowan process was higher than that in the case without SFTS (Fig. \ref{V5_NoSFTS}) and the CRSS for the 60\degree configuration with SFTS $\tensor\epsilon^{0}_3$ was smaller than that obtained in the absence of the SFTS (Fig. \ref{V5_NoSFTS}).

In the case of  $\tensor\epsilon^{0}_5$ (Fig. \ref{V5_SFTS}c), the dislocation line is initially repulsed by the precipitate (region i) but it is strongly attracted afterwards towards the broad face of the precipitate (region ii). The final Orowan loop is in contact with the precipitate along the whole matrix/precipitate interface (regions iii and iv).  This leads to a very small radius of curvature of the dislocation and the CRSS in this case is much higher than the one in the absence of the SFTS (Fig. \ref{V5_NoSFTS}). The situations in the presence of the two other SFTS (Fig. \ref{V5_SFTS}b and d) are in between those reported above and the CRSS in these cases were  equal  ( $\tensor\epsilon^{0}_4$)  or slightly higher ($\tensor\epsilon^{0}_6$) than that in the absence of the SFTS.

\subsection{Influence of the aspect ratio of the precipitates}

Although $\theta'$ precipitates in Al-Cu alloy have a large aspect ratio, this geometric feature may be changed by the addition of alloying elements which modify the interfacial energy between the Al matrix and the precipitate \citep{MMR00, YZS16, DWG17}. Thus, it is interesting to analyze the influence of the precipitate aspect ratio on the mechanisms of dislocation-precipitate interaction in the presence of the SFTS. To this end, DDD simulations in the 0\degree and 60\degree configuration were carried out with precipitates with  aspect ratios  in the range 26:1 to 1:1 while the precipitate  volume fraction (3.1 10$^{-4}$) was held constant. 

\begin{figure}
\centering
\includegraphics[width=0.49\textwidth]{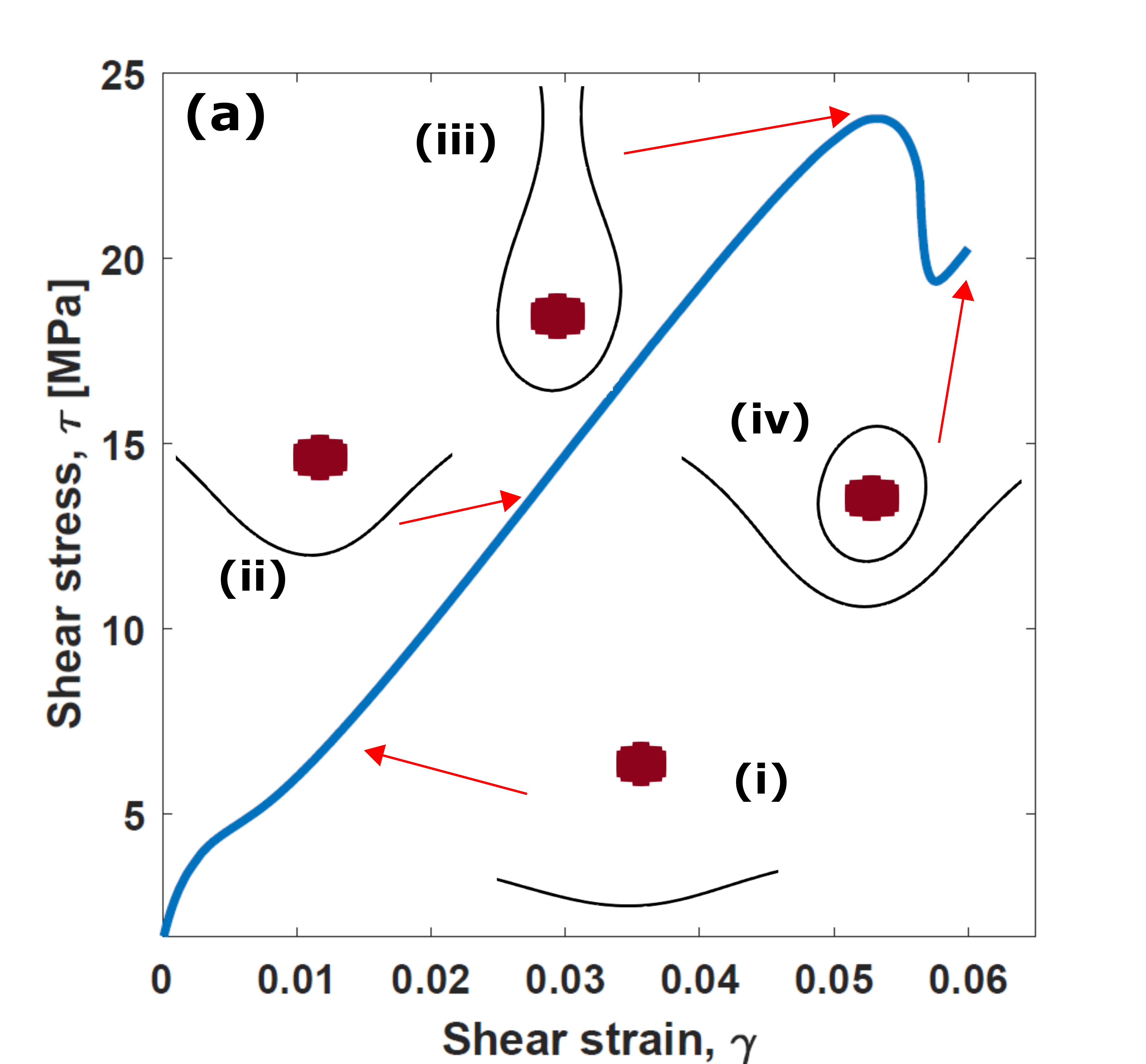}
\includegraphics[width=0.49\textwidth]{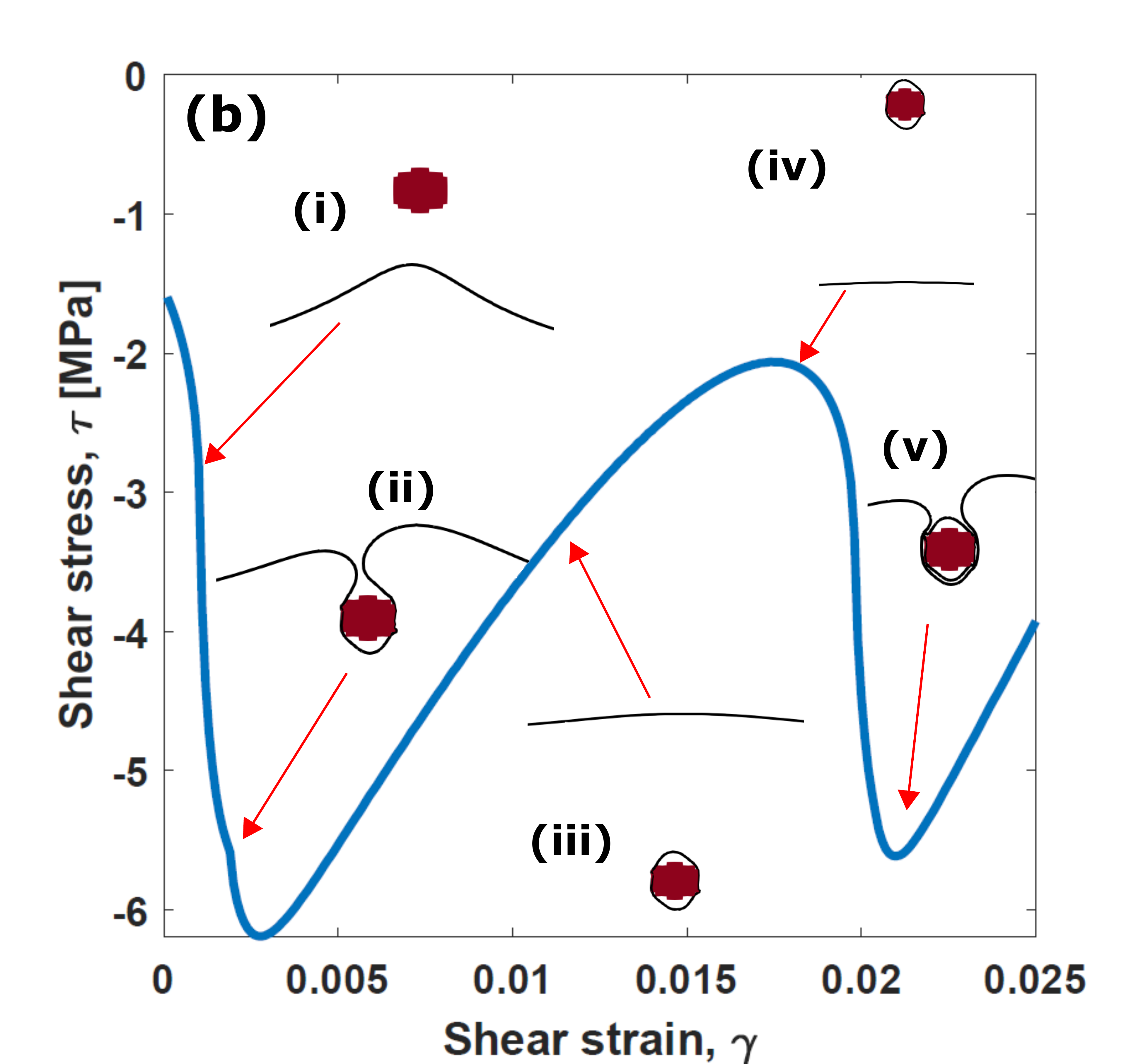}
\caption{Shear stress-strain curve corresponding to the dislocation/precipitate interaction in the 0\degree configuration with SFTS. (a) $\tensor\epsilon^{0}$ = $\tensor\epsilon^{0}_1$. (b) $\tensor\epsilon^{0}$ = $\tensor\epsilon^{0}_2$. The evolution of the dislocation line around the precipitate is showed in several points along the curve.}  
\label{V1_AR1_SFTS}
\end{figure}

\begin{figure}[!]
\centering
\includegraphics[width=\textwidth]{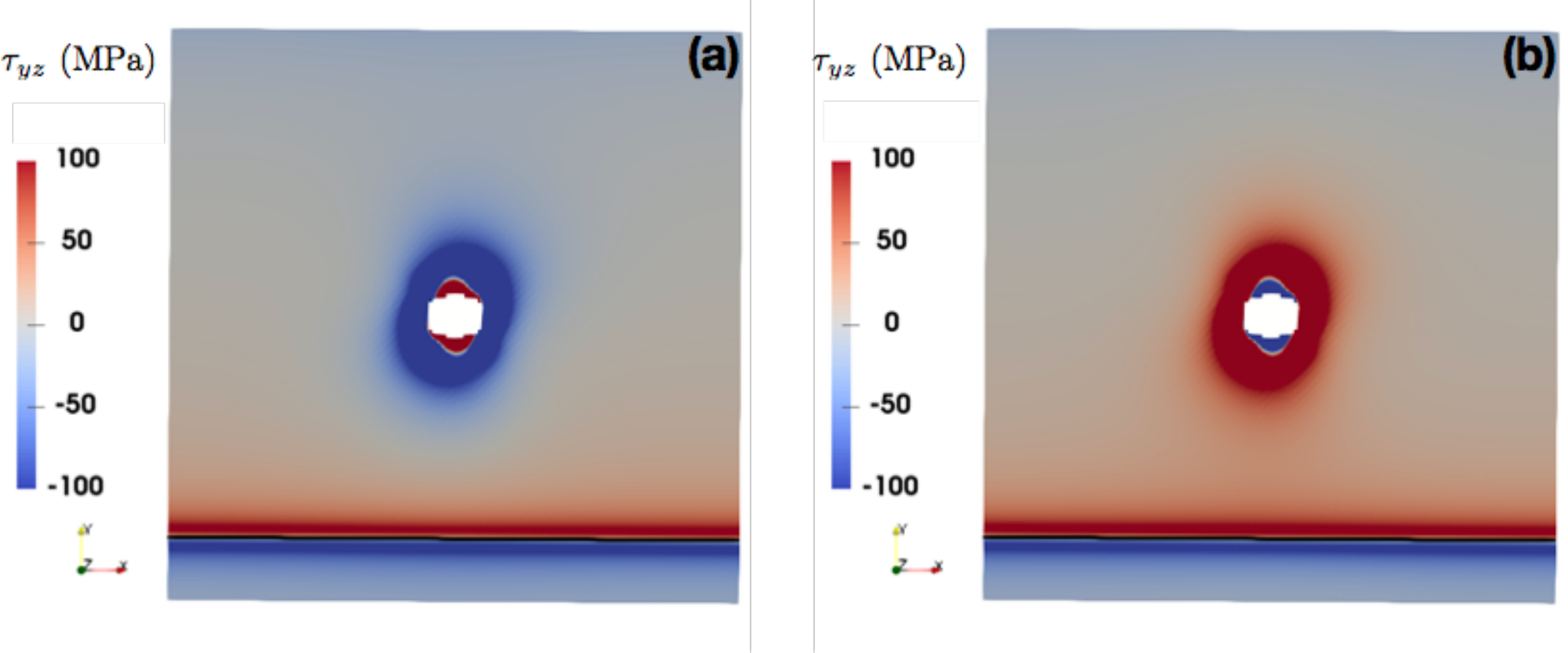}
\caption{Contour plots of the shear stress, $\tau_{yz}$ corresponding to the dislocation/precipitate interaction in the 0\degree configuration. (a) SFTS $\tensor\epsilon^{0}$ = $\tensor\epsilon^{0}_1$.  (b) SFTS $\tensor\epsilon^{0}$ = $\tensor\epsilon^{0}_1$. The dislocation is shown as a black line and the cross-section of the precipitate is white.}  
\label{Stress-AR1}
\end{figure}

In the absence of SFTS, the dislocations overcome the precipitate by the formation of an Orowan loop and the corresponding results are not plotted for the sake of brevity. The shear stress-strain curves corresponding to the 0\degree configuration with SFTS given by $\tensor\epsilon^{0}_1$ and $\tensor\epsilon^{0}_2$ are plotted in Figs. 
\ref{V1_AR1_SFTS}a) and b), respectively.  The corresponding contour plots of the shear stress $\tau_{yz}$ in the initial configuration are shown in Figs. \ref{Stress-AR1}a) and b).

In the first case (\ref{Stress-AR1}a), the stress field induced by the SFTS repels the dislocation (region i) and impedes that the dislocation line gets in touch with the precipitate (region ii). As a result, the effective precipitate diameter that controls the radius of curvature of the dislocation arms to form an Orowan loop is increased (region iii). On the contrary, the stress field created by $\tensor\epsilon^{0}_2$ (Fig. \ref{V1_AR1_SFTS}b) strongly attracts the dislocation line (regions i and ii) and  the dislocation spontaneously overcomes the precipitate by the formation of a very tight Orowan loop (region ii). This process is repeated is further strain applied to the simulation domain (regions iv and v).

\begin{figure}[t]
\centering
\includegraphics[width=0.48\textwidth]{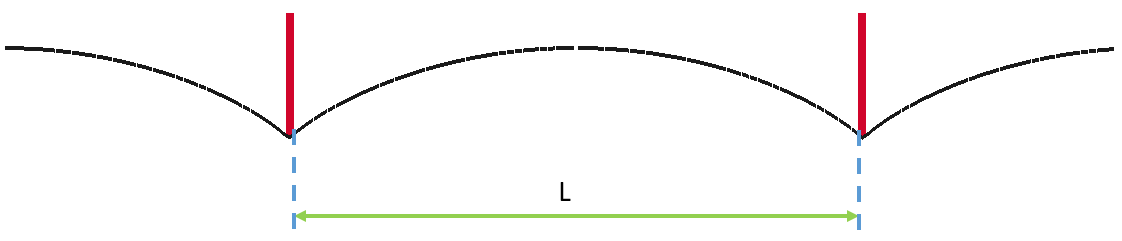}
\includegraphics[width=0.48\textwidth]{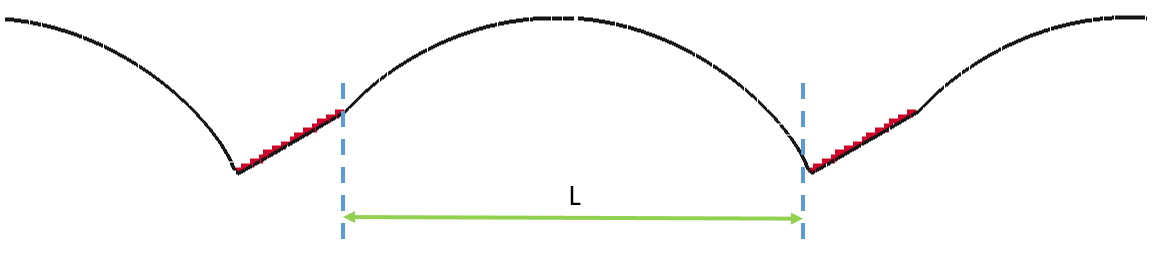}
\includegraphics[width=\textwidth]{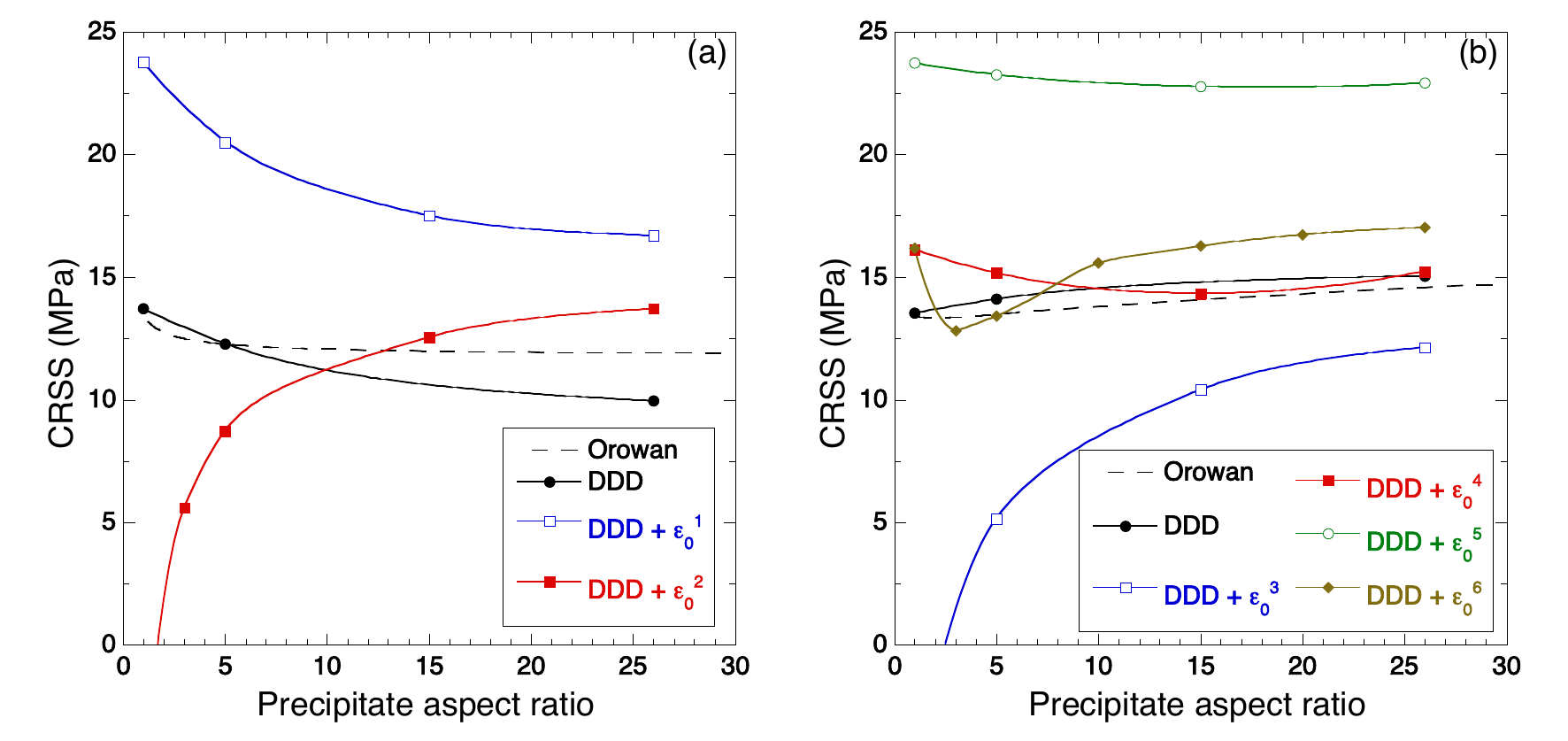}
\caption{CRSS as a function of the precipitate aspect ratio. (a) Precipitate in the 0\degree configuration. (b) Precipitate in the 60\degree configuration. The results corresponding to DDD with and without SFTS as well as to the Orowan model are presented in each figure for comparison.}  
\label{CRSS}
\end{figure}

The mechanisms of dislocation-precipitate interaction in the case of 60\degree con\-figuration with smaller aspect ratio are qualitatively similar to those reported above and are not included for the sake of brevity. It is, however, important to assess the influence of the SFTS and of the precipitate aspect ratio on the CRSS for both precipitate orientations. These results are plotted in Figs. \ref{CRSS}a) and b) for the 0\degree and 60\degree orientations, respectively. The CRSS obtained from the simulations with and without SFTS are included in each figure, together with the predictions of the Orowan model for the CRSS, $\tau_O$, which is given by 

\begin{equation}
\tau_O = \frac{Gb}{L}
\label{Orowan}
\end{equation}

\noindent where $G$ (= 29.9 GPa) is the shear modulus of the Al matrix in the slip plane parallel to the Burgers vector $b$ (= 0.2863 nm) and $L$ is the distance between precipitates along the dislocation line (Fig. \ref{CRSS}). In the case of the 0\degree orientation, the variation of the precipitate aspect ratio from 26:1 to 1:1 (while the precipitate volume fraction was held constant) did not change $L$ significantly (from $L$ = 718 nm to 636 nm, respectively), while the differences in $L$ with the aspect ratio were slightly different in the 60\degree configuration (from $L$ = 587 nm  for 26:1 to $L$ =  637 nm for 1:1 aspect ratio). Thus, the CRSS given by eq. \eqref{Orowan} was almost constant for the 0\degree configuration and decreased slightly with the aspect ratio in the 60\degree configuration (Fig. \ref{CRSS}). The predictions of the Orowan model were in good agreement with the DDD simulations in the absence of the SFTS in the 60\degree configuration but they overestimated by $\approx$ 20\% the CRSS for precipitates with large aspect ratio oriented at 0\degree. These results are in agreement with the main hypotheses of the Orowan model, which assumed that the precipitates were spherical (small aspect ratio) and that the dislocation line formed a circular loop between precipitates.

The introduction of the SFTS led to large variations in the CRSS, which were more important if the precipitates had small aspect ratio. Depending on whether the stress field associated with the SFTS attracted or repelled the dislocation, the CRSS could increase or decrease dramatically in the precipitates in the 0\degree orientation (Fig. \ref{CRSS}a) when the aspect ratio was 1:1. As shown in Fig.  \ref{V1_AR1_SFTS}, the stress field associated to the SFTS controlled the shape of the dislocation loop at the instability point and, thus, the magnitude of the CRSS. Similar results were obtained in the case of the precipitate oriented at 60\degree with an aspect ratio of 1:1 (Fig. \ref{CRSS}b) for $\tensor\epsilon^{0}_5$ and $\tensor\epsilon^{0}_3$. It should be noted, however, that the SFTS is an important factor to determine the shape of the precipitate, as shown by \cite{LBL17}. The SFTS used in this simulations led to precipitates with large aspect ratio (> 10) and may not be realistic for equiaxed precipitates. 

The influence of the SFTS decreased as the precipitate aspect ratio increased because the dislocation loop configuration depended not only in the SFTS but also on the precipitate shape, leading always to an elongated loop parallel to the main axis of the precipitate (Figs. \ref{V1_SFTS} and \ref{V5_SFTS}).  Nevertheless, it should be noticed that the presence of the SFTS increased the CRSS of precipitates with an aspect ratio of 26:1 in all cases (with the exception of the SFTS $\tensor\epsilon^{0}_3$ in the 60\degree configuration) the CRSS with respect to the values obtained by DDD simulations without the SFTS. These results indicate that the stress fields around the precipitate (due to the SFTS or to thermal stresses generated upon cooling from the ageing temperature as a result of the thermal expansion mismatch between the matrix and the precipitates) have to be taken into account to make quantitative predictions of the strengthening provided by precipitates in metallic alloys. It should be finally noted that the presence of the SFTS $\tensor\epsilon^{0}_6$ in the 60\degree configuration changed the shape of the Orowan loop around the precipitate when the aspect ratio increased from 1:1 to 2:1 and again for larger values of the aspect ratio, leading to a complex variation of the CRSS with this parameter for this particular SFTS. Moreover, the stress fields around precipitates can interact with each other for large precipitate volume fractions (far away from the dilute conditions of this investigation), leading to complex interaction patterns between dislocations and precipitates.

\subsection{Influence of the elastic mismatch between matrix and precipitate}

\begin{figure}[t]
\centering
\includegraphics[width=\textwidth]{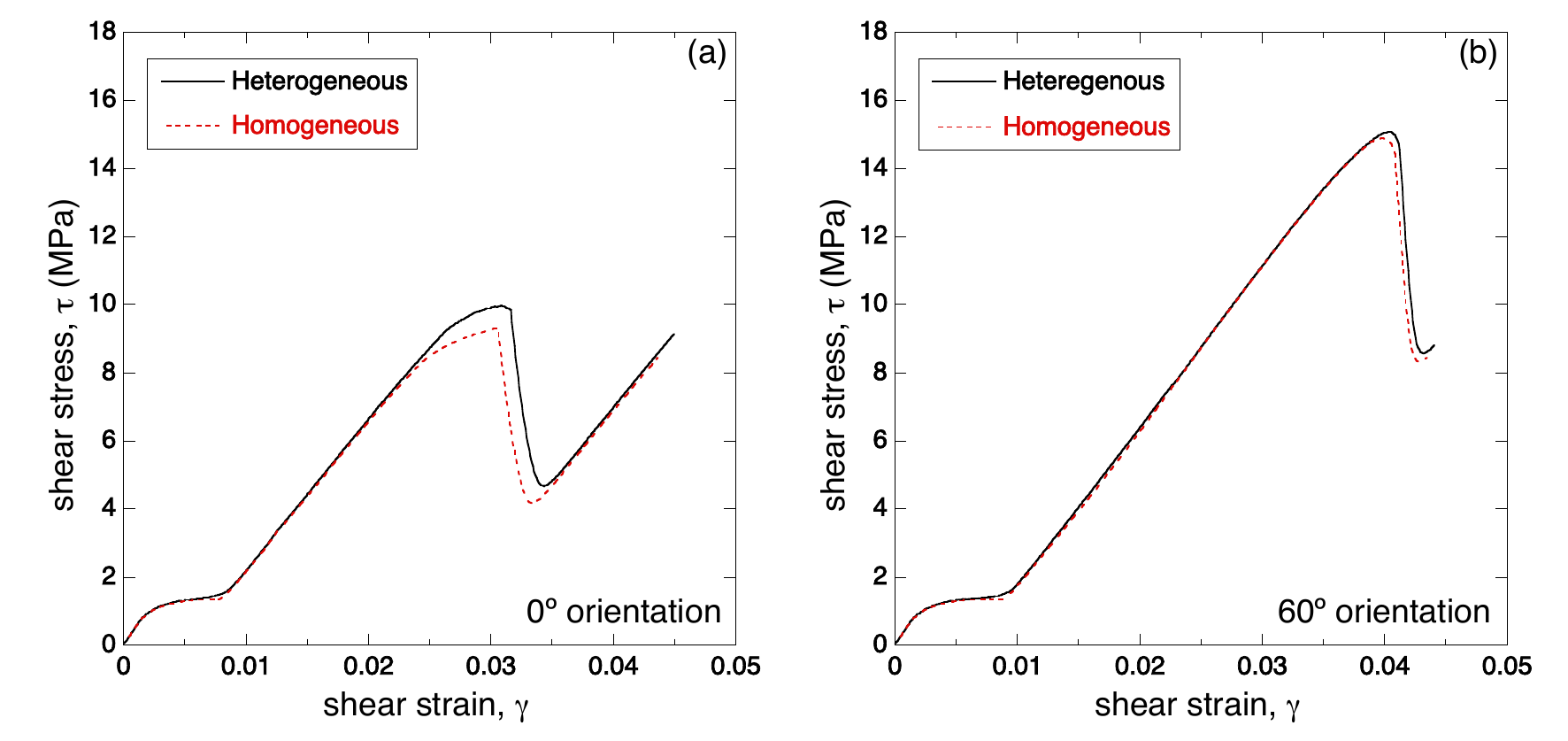}
\caption{Shear stress-strain curves corresponding to the dislocation/precipitate interaction with homogeneous and heterogeneous elastic constants for the matrix and the precipitate. (a) 0 \degree configuration without SFTS. (b) 60\degree configuration without SFTS.}  
\label{HH}
\end{figure}

All the results presented above were obtained using different values of the elastic properties for the matrix and the precipitate, according to the DFT results in Tables 1 and 2. However, it is interesting to assess the influence of the elastic heterogeneity on the stress-strain curves and on the CRSS. Thus, two simulations were carried out for the 0\degree and 60\degree orientations (without STFS) for precipitates with an aspect ratio 26:1 in which the elastic properties of the precipitate were identical to those of the matrix. The corresponding stress-strain curves for these homogeneous simulations are plotted in Figs. \ref{HH}a) and b) for the precipitates oriented at 0\degree and 60degree, together with the results obtained with the actual elastic constants of the matrix and the precipitate. The differences in the mechanisms and in the CRSS were negligible, in agreement with previous investigations  \citep{Shin2003_DDD_FEM}.

\section{Concluding Remarks}

\label{sec:Conclusions}

The mechanisms of dislocation/precipitate interaction were analyzed by means of  a novel DDD strategy  based in the DCM \citep{Lemarchand2001_DCM} in combination with a FFT solver to compute the mechanical fields \citep{Capolungo2015_DDDFFT}.  This framework neither requires the use of analytical expressions for the displacement fields of the dislocation segments (and, thus, can be easily extrapolated to anisotropic materials), nor the computational power increases with the square of the number of dislocations segments. Moreover, very fine discretizations (necessary to model precipitates with large aspect ratio) can be used owing to the efficiency of the FFT solver, and  the influence of the image stresses (induced by the elastic modulus mismatch between the matrix and the precipitate) and of the SFTS can be easily incorporated to the simulations. The original DDD strategy \citep{Capolungo2015_DDDFFT} was modified to include straight dislocation segments by means of the FDM model, the appropriate configuration to analyze the interaction of a single dislocation line with a precipitate.

The novel DDD model was used to analyze the mechanisms of dislocation/precipitate interaction and the corresponding CRSS in Al-Cu alloys. The orientation, size and shape of the $\theta'$ precipitates as well as the SFTS associated to the different precipitate variants were obtained from recent multiscale modelling simulations based on the phase-field model \citep{LBL17}, while the elastic constants of the Al matrix and of the precipitates were calculated by DFT and the dislocation mobility as a function of the dislocation character was obtained from molecular dynamics simulations \citep{Molinari2017_Mobility}. This leads to a multiscale modeling strategy, in which all the parameters in the DDD simulations are obtained from calculations at lower length scales.

The DDD simulations provided for the first time a detailed account of the influence of the precipitate aspect ratio, orientation, SFTS and elastic mismatch between the matrix and the precipitate on the dislocation path to form an Orowan loop and on the CRSS to overcome the precipitate. It was found than the elastic mismatch have a negligible influence on the dislocation/precipitate interaction in the Al-Cu system while the influence of the precipitate aspect ratio and orientation was reasonably captured by the simple Orowan model in the absence of the SFTS. Nevertheless, the introduction of the SFTS led to dramatic changes in the dislocation/precipitate interaction and in the CRSS. This effect decreased as the precipitate aspect ratio increased but it was still very important (above 50\% in the CRSS for some precipitate variants) for $\theta$' precipitates with the typical aspect ratio found in Al-Cu alloys. Thus, this investigation reveals the large influence of the SFTS on the mechanics of dislocation/precipitate interaction, an important factor that has not been previously considered in the analysis of precipitation hardening.

Finally, The methodology presented in this paper opens the possibility to explore in more detail the mechanisms of dislocation/precipitate interaction in metallic alloys with realistic values of the precipitate size, shape and aspect ratio as well as of the elastic mismatch and of the dislocation mobility. They will be able to provide quantitative assessments of the strengthening provided by the precipitates, taking into account the influence of the SFTS and of the thermal stresses that develop upon cooling from the ageing treatments at high temperature. Finally, they can be extended to deal with larger volume fraction of precipitates to account for the interaction between the SFTS of different precipitates  and to model the propagation of a dislocation through a forest of precipitates including the effect of cross-slip. These topics will be the subject of future investigations.

\section{Acknowledgments}

This investigation was supported by the European Research Council under the European Union's Horizon 2020 research and innovation programme (Advanced Grant VIRMETAL, grant agreement No. 669141). LC was funded by the US Department of EnergyÕs Nuclear Energy Advanced Modeling and Simulation (NEAMS).

\appendix
\section{Influence of strain rate}
\begin{figure}[h]
\includegraphics[width=1\textwidth]{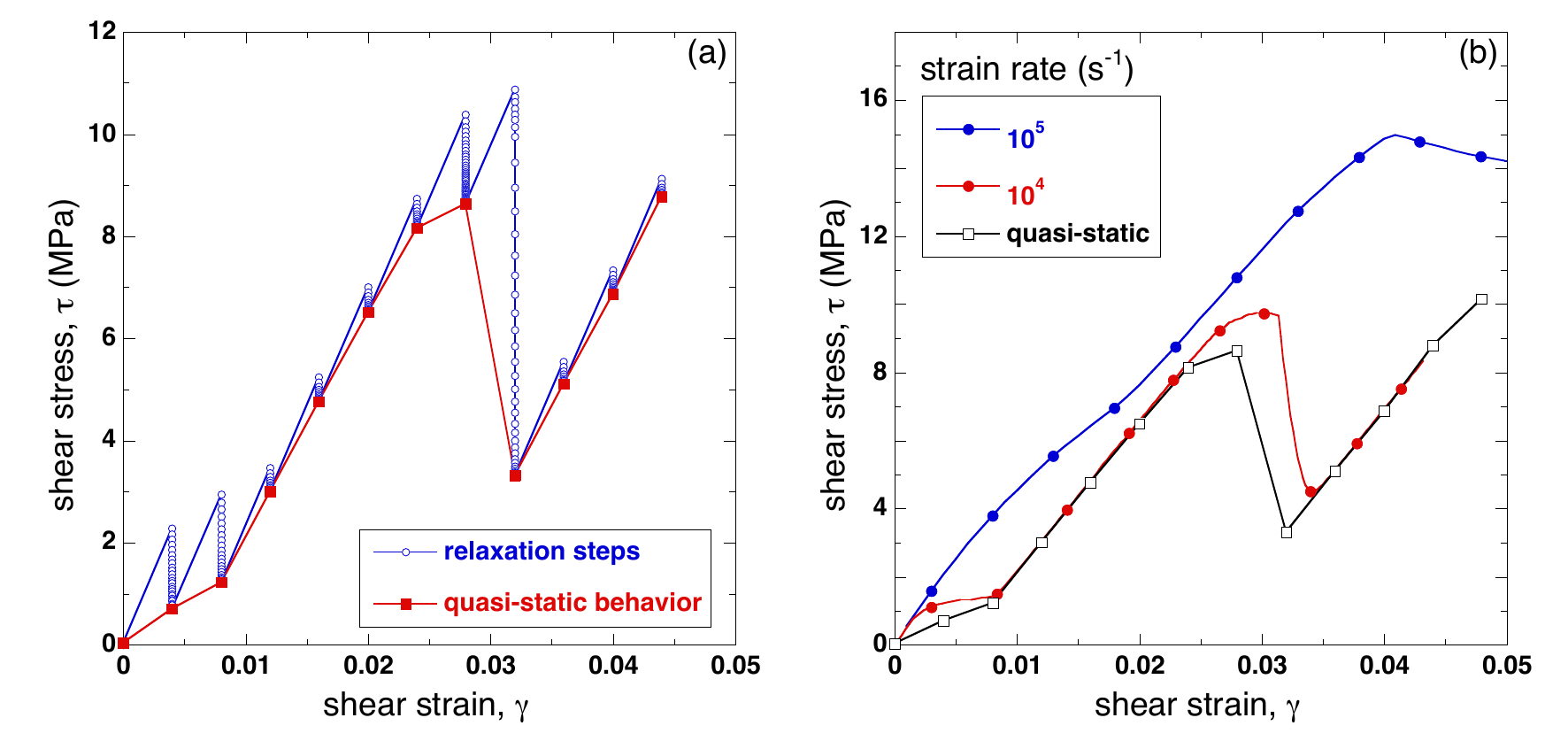}
\caption{(a) Shear stress-strain curve of the dislocation precipitate interaction obtained using the relaxation process. (b) Comparison between shear stress-strain curves of the dislocation/precipitate interaction as a function of the applied strain rate. The quasi-static results  correspond to the red curve in (a). See text for details.}  
\label{qstatic}
\end{figure}

Dislocation dynamics simulations are normally carried out at high strain rates (10$^2$ to 10$^5$ s$^{-1}$) for computational reasons and this limitation often leads to question whether the results obtained are applicable under quasi-static conditions.  In order to analyze this effect,  several simulations were carried out using a relaxation strategy that allows to study the dislocation dynamics under quasi-static conditions. 
In this approach, a strain increment is applied to the simulation box at at a high strain rate, in this case 4.0 10$^{7}$ s$^{-1}$, and the energy of the system is relaxed afterwards  during several steps at a constant applied strain. The shear stress is reduced during relaxation and the process is finished when the difference in the shear stress between to consecutive relaxation steps is lower than a certain tolerance, and the system can be considered to be in equilibrium. Then, a new strain increment is applied and the whole relaxation process is repeated. The shear stress-strain curve obtained following this process is plotted in Fig. \ref{qstatic}a) in the case of the interaction of an edge dislocation with a precipitate with a diameter of 156 nm and an aspect ratio of 26:1 in the 0\degree configuration. The blue line with open symbols shows the successive strain increments followed by the relaxation of the shear stress and the red line with solid symbols stands for the quasi-static shear stress-strain curve. The shear stress-strain curves obtained at different strain rates (10$^4$ and 10$^5$ s$^{-1}$) for this case are plotted in Fig. \ref{qstatic}b), together with the quasi-static curve in Fig. \ref{qstatic}a). The comparison between these curves shows that the results obtained at an applied strain rate of 10$^4$ s$^{-1}$ were very close to the quasi-static simulations and, thus, the DDD presented in this paper were carried out at an applied strain rate of 10$^4$ s$^{-1}$.\\

\end{document}